\documentclass[10pt]{article}
\usepackage{amsmath,amsopn,amscd,amsthm,amssymb,xypic,rotating,epic}

\xyoption{all}

\newcommand{\double}       {\baselineskip 12pt}
\newcommand{\single}       {\baselineskip 12pt}
\newcommand{\dqt}[1]        {``{#1}"}

\newtheorem{theorem}{Theorem}[section]

\newcommand{\cmd}[1]      {\underline{{#1}}}
\newcommand{\cb}          {\begin{tabbing}MMMMM\=MM\=MM\=MM\=MM\=MM\=MM\=MM\=MM\=MM\= \kill}
\newcommand{\ce}          {\end{tabbing}}

\newcommand{\separate}     {\vspace{0.3cm}\begin{center}*~~~~~~~~~~*~~~~~~~~~~*\end{center}\vspace{0.3cm}}

\def\emph{\textsl}
\def\em{\sl}
\def\textbf{\pmb}
\def\dfeq{\stackrel{\triangle}{=}}

\newcounter{myremark}
\newcounter{myexample}
\def\example{
\bigskip

\refstepcounter{myexample}%
\noindent \textbf{Example \Roman{myexample}:}\\
}

\begin{document}

\title{Hiding from Facebook: An Encryption Protocol resistant to Correlation Attacks}
\author{Chen-Da Liu and Simone Santini}
\date{Universidad Aut\'onoma de Madrid}

\maketitle

\double

\begin{abstract}
  In many social networks, one publishes information that one wants to
  reveal (e.g., the photograph of some friends) together with
  information that may lead to privacy breaches (e.g., the name of
  these people). One might want to hide this sensitive information by
  encrypring it and sharing the decryption key only with trusted
  people, but this might not be enough. If the cipher associated to a
  face is always the same, correlation between the output of a face
  recognition system and the cipher can give useful clues and help
  train recognitors to identify untagged instances of the face. We
  refer to these as \emph{correlation attacks}.

  In this paper we present a coding system that attempts to counter
  correlation attacks by associating to each instance of a face a
  different encryption of the same tag in such a way that the
  correlation between different instances is minimal.

  In addition, we present a key distribution code that allows only the
  owner of the images to encode the tags, but allows a group of
  trusted friends to decode them.
\end{abstract}

\section{Introduction}
Privacy and identity protection on the Internet is today an issue of
primary importance, one of those issues that escape the confines of
the technical literature and trickles into the mainstream media as
well as in the legal profession \cite{belanger:11}. In the technical
field, a considerable interest is directed towards protecting our
private data from malicious and unauthorized third parties \cite{chen:12} but
with mounting pressure from governments and data-hungry corporations,
even exposing data to the company that hosts them has become a
potential source of privacy breaches \cite{young:13}.

One of the most powerful tools for companies to discover facts about
us, much beyond the data that we believe we gave them, is the mutual
information between different data \cite{batina:11}. As an example of
the things that we can derive from apparently harmless information, a
year 2000 article \cite{santini:00e} considered snapshots taken by web
cameras at several spots on the freeways of central Seattle
(WA). Based on these data, and using Network Tomography
\cite{vardi:96}, it was possible to determine the major paths followed
by commuters in the city. It was possible, for example, to determine
which residential areas where white-collar based on the areas in which
its dwellers predominantly worked, and the predominant work schedule.

In the motivating example of this work, somebody publishes a lot of
pictures of friends and family together with tags containing the name
of the people that appear in then. The collection of tagged pictures
and the association that can be inferred between people that appear
together in them allow the host of the social network to statistically
deduce a significant amount of information, potentially infringing the
privacy of the people involved.

The algorithms for face recognition in crowded scenes are imperfect
\cite{zafeiriou:15}, but the presence of the tags greatly helps them
(thus endangering privacy) by identifying people without need of face
recognition, and by providing a tagged data set on which more reliable
recognition algorithms can be trained, improving the identification of
the same people in images in which they are not tagged. Note that many
social networks do provide means to hide content from other users, but
these methods are ineffective if the unwanted access is carried out by
the owner of the social network.

In our scenario, encrypting the tags with a standard method would be
of little help: although tags would be encoded, the name of the same
person would have the same code in all the pictures so there would be
a consistent relation between the faces and the encrypted tags, and
the correlation between codes and images would be preserved. Moreover,
with a recognition algorithm trained on a lot of encrypted (but
identical) examples, the presence of one unencrypted tag would be
enough to reveal the identity of the person in all encrypted pictures,
thus providing a \dqt{ground truth} that will help code breaking.

Based on this motivating example, in this paper we develop a method to
reduce the mutual information between the \emph{manifest content} of a
message (in our example: the images of people that are openly
published without encryption) and the \emph{hidden content} (the tag
with the name of the people that appear in the pictures). In a
nutshell, the system works in such a way that each instance of the tag
is translated into a different cipher. All these ciphers can be
decoded to recover the (unique) originao tag using the same key, and
are created in such a wat that the correlation between them is minimal. 

The encryption, in our scenario, can't be done using public key
encryption methods \cite{rivest:78}; in public key encryption, the
encryption key is public and freely distributed, so that everybody may
encrypt a message, while the decryption key is kept private, so that
only a person (or a trusted group) may decode. On the other hand, in
our case, only the person who published the tags must be able to
encode it, while only the members of a trusted group must be able to
decode it. Therefore, we need a scheme that allows a person to share a
decoding key with a trusted group without revealing it to a malicious
agent observing the traffic.

\section{The Model}
The data that we consider are pairs of messages $(m_i,h_i)$; $m$ is
the \emph{manifest content}, data that we publish as they are and that
we allow everybody to see (the image of a face, in our example); $h$
is the \emph{hidden content}, data about $m$ that we don't want to
reveal (the name of the person). Many pairs of such data will be
published. The $m_i$ are not necessarily identical, but there is an
algorithm that applied to $m_i$ and a prorotype $m$, gives us the
probability that $m_i$ is the same as $m$:
$p_m(m_i)={\mathbb{P}}\{m_i=m\}$. If there is a correlation between
$m$ and $h$, and if $h$ can be detected with higher precision than
$m$, this will reduce the uncertainty on the detection of $m$, thus
providing information that, due to uncertainty, would not be available
from $m$ alone. We call \emph{correlation attacks} the attempts to
infer information based on the correlation between $m_i$ and $h_i$. 

\example
We present a simple example to illustrate our point. Let $(m,h)$ be
the manifest and hidden content and assume that the ground truth for
them is $m=h=0$. Assume that the \dqt{hidden} content is, in this
case, not quite hidden, and that it can easily be estimated.  We
measure $m$ and, assuming that we have an unbiased estimator with a
Gaussian error, we have a probability distribution for the measurement
\begin{equation}
  p_m(x) = \frac{1}{\sqrt{2\pi{\sigma_m^2}}} \exp{\left(-\frac{x^2}{\sigma_m^2}\right)}
\end{equation}
where $\sigma_m^2$ is the variance of the measurement error. The data
$h$ are also measured with a Gaussian error, but with a variance
$\sigma_h<\sigma_m$:
\begin{equation}
  p_h(x) = \frac{1}{\sqrt{2\pi{\sigma_h^2}}} \exp{\left(-\frac{x^2}{\sigma_h^2}\right)}
\end{equation}
Assume that between the two there is a positive correlation
$0<\rho\le{1}$. We can use $h$ to estimate $m$ by writing
$p(m)=p(m,h)/p(h)$. In order to do this, we need to model the relation
between $m$ and $h$. Here we shall assume, for the sake of
exemplification, that it is linear, that is, $m=\beta{h}$. Also,
define $\sigma=\beta\sigma_h$. Then we can write
\begin{equation}
  \begin{aligned}
    \log p(m,h) & \sim -\frac{1}{2(1-\rho^2)} \left[ \frac{1}{\sigma_m^2} - 
                                                     \frac{2\rho}{\beta\sigma_m\sigma_h} + 
                                                     \frac{1}{\beta^2\sigma_h^2} 
                                            \right] m^2 \\ 
    &= -\frac{1}{2(1-\rho^2)} \left[ \frac{1}{\sigma_m^2} - 
                                     \frac{2\rho}{\sigma\sigma_h} + 
                                     \frac{1}{\sigma^2} 
                             \right] m^2
  \end{aligned}
\end{equation}
That is,
\begin{equation}
  \begin{aligned}
    \log p(m) &= \log p(m,h) - \log p(h) \\
    &\sim -\frac{1}{1-\rho^2} \left[ \frac{1}{2\sigma_m^2} - \frac{\rho}{\sigma\sigma_m}
                                     + \frac{-\frac{1}{2} + \rho^2}{\sigma^2} \right] m^2
  \end{aligned}
\end{equation}
The term in parenthesis is greater than $1/\sigma_m^2$ if
\begin{equation}
  \rho > \frac{1}{\sqrt{2}}+\frac{1}{2} \frac{\sigma\sigma_m}{\sigma^2+\sigma_m^2}
\end{equation}
if $h$ can be detected without uncertainty ($\sigma_h=0$), then the
uncertainty on the detection of $m$ is reduced if $\rho>1/\sqrt{2}$.

\separate

From the previous example we see that in order to reduce the amount of
information that we can extract from $m$, we have either to make $m$
harder to detect, increasing $\sigma_m$, or reduce $\rho$, the
correlation between $m$ and $h$. The value of $\sigma_m$ is usually
not under the control of the publisher: in the case of images, it
depends on the quality of the processing algorithms available to the
intruder. The most obvious way to protect $m$ is to reduce the
correlation with the easy-to-detect tag $h$.

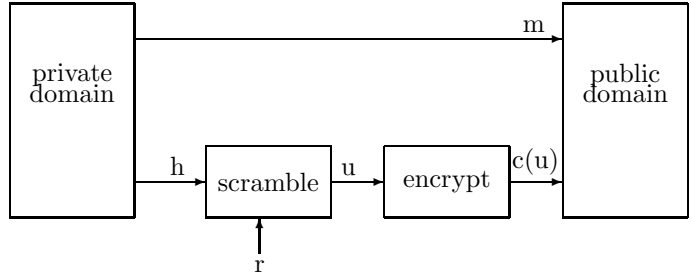
\begin{figure}
  \begin{center}
    \setlength{\unitlength}{1.35em}
    \begin{picture}(20,7)(0,-1)
      \multiput(0,0)(15.5,0){2}{
        \multiput(0,0)(3.5,0){2}{\line(0,1){6}}
        \multiput(0,0)(0,6){2}{\line(1,0){3.5}}
      }
      \put(-0.5,0) {
        \multiput(6,0)(5,0){2}{
          \multiput(0,0)(3.5,0){2}{\line(0,1){2}}
          \multiput(0,0)(0,2){2}{\line(1,0){3.5}}
          \put(3.5,1){\vector(1,0){1.5}}
        }
        \put(4,1){\vector(1,0){2}}
        \put(4,5){\vector(1,0){12}}
        \put(7.5,-1){\vector(0,1){1}}
        \put(15.5,5.2){\makebox(0,0)[rb]{m}}
        \put(15.9,1.2){\makebox(0,0)[rb]{c(u)}}
        \put(10,1.2){\makebox(0,0)[b]{u}}
        \put(5,1.2){\makebox(0,0)[lb]{h}}
        \put(7.5,-1.2){\makebox(0,0)[t]{r}}
        \put(7.75,1){\makebox(0,0){scramble}}
        \put(12.75,1){\makebox(0,0){encrypt}}
      }
      \put(1.75,4){\makebox(0,0){private}}
      \put(1.75,3.5){\makebox(0,0){domain}}
      \put(17.25,4){\makebox(0,0){public}}
      \put(17.25,3.5){\makebox(0,0){domain}}
    \end{picture}
  \end{center}
  \caption{The general encryption schema. The manifest content $m$ is
    passed as is to the public domain. The hidden content $h$ is first
    scrambled with the aid of a random number $r$ so that different
    copies of the same manifest content $m$ will be associated to
    different, and uncorrelated, contents $u$. The content $u$ is then
    encrypted to obtain the code $c(u)$, which is released in the
    public domain.}
  \label{general}
\end{figure}

The general schema of our encryption strategy is shown in
Figure~\ref{general}.
The encryption of the hidden content $h$ is carried out in two steps:
the first one is a randomization, in which $h$ is transformed into a
message $u$ with the intervention of a random number $r$. The
randomization must be reversible independently of $r$, that is, if
$u=f(h,r)$, there must be a function $g$ such that, for all $h,r$,
$g(f(h,r))=h$, and $g$ is computable. At the same time, the
correlation between $h$ and $f(h,r)$ must be as little as possible.

The second step is a standard encryption method that, given $u$,
produces the cipher $c(u)$ which is released into the public domain
associated to the manifest content $m$. As we mentioned, the nature of
the problem (allowing a trusted group of people to decode $c(u)$)
prevents us from using a public key method, so we shall use a
symmetric encryption and devise a safe method for the distribution of
the key \cite{mahalanobis:08}.

\section{Generation of the uncorrelated message $u$}
The purpose of this section is to generate $u$ as a function of $h$
and a random number $r$ in such a way that $h$ can be recovered from
$u$ but such that a statistical analysis of any number of instances of
$u$s derived from the same $h$ will not reveal information about
$h$. We assume that the message to be encoded is an integer
number. This assumption does not limit the genericity of our method
since, clearly, anything that can be stored in a computer memory may
be seen as an integer number. 

That is, we want to determine a function
$f:{\mathbb{N}}^2\rightarrow{\mathbb{N}}$ such that $u=f(h,r)$, where
$r$ is a random number such that:
\begin{description}
\item[i)] there is a computable function $g$ such that, for all $h$
  and $r$, $h=g(f(h,r))$;
\item[ii)] Given any number of pairs $(h,f(f,r_i))$ for different
  $r_i$, no statistical analysis will allow us to predict $h$ with
  significant certainty.
\end{description}
As a statistical test, we choose Canonical Correlation Analysis (CCA,
\cite{hardle:15}). The reason for this is that CCA doesn't simply
measure the statistical relation between $h$ and $u$, but seeks a
(linear) transformation that maximizes this relation.

\example 
Let $h$ be represented using $b$ bits, and let $u$ be
represented using $B>b$ bits. We can obtain a function $f$ which
satisfies i) by placing the $b$ bits of $h$ in fixed position and
filling in the remaining $B-b$ with random bits. Knowing the position
of the bits of $h$ allows us to define the function $g$. The mutual
information $I(h,u)$ can be made as small as desired by increasing
$B$, but, given a sufficient number of pairs, CCA will easily discover
the projection on the $b$ \dqt{right} bits, a projection that will
give a correlation of $1$.

\separate

Given the weakness of placing random bits in fixed positions, we use a
different method, more robust to this kind of attacks. Consider the
numbers
\begin{equation}
  p_n = \frac{n(n+1)}{2}
\end{equation}
for which the relation $p_n-p_{n-1}=n$ holds. Between the numbers
$p_n$ and $p_{n+1}$ we can imagine $n$ free \dqt{slots}, beginning
with $p_n$ and ending with $p_{n+1}-1$, in which we can accommodate
codes for the numbers $0,\ldots,n-1$. Each number $p_m$, $m>n$ will
also have, between $p_m$ and $p_{m+1}$ the same $n$ slots (it will
actually have $m$), so given a number $n$ we can encode it as the
$n$th number after any of the $p_m$ with $m\ge{n}$. Therefore, given a
number $n$, we choose a random number $r\ge{n}$ and we encode $n$ as
$p_r+n$. This code will always be in the interval $[p_r,p_{r+1})$.

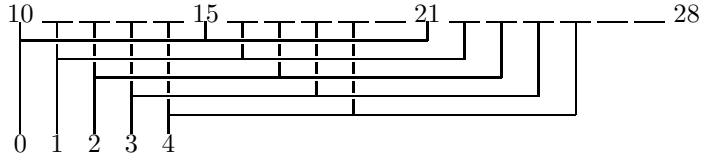
\begin{figure}
  \begin{center}
  \setlength{\unitlength}{0.7em}
  \begin{picture}(36,8)(0,-1)
    \put(0,6){\makebox(0,0)[b]{10}}
    \put(0,6){
      \multiput(2,0)(2,0){4}{\put(-0.8,0){\line(1,0){1.6}}}
    }
    \put(10,6){\makebox(0,0)[b]{15}}
    \put(10,6){
      \multiput(2,0)(2,0){5}{\put(-0.8,0){\line(1,0){1.6}}}
    }
    \put(22,6){\makebox(0,0)[b]{21}}
    \put(22,6){
      \multiput(2,0)(2,0){6}{\put(-0.8,0){\line(1,0){1.6}}}
    }
    \put(36,6){\makebox(0,0)[b]{28}}
    \put(0,0){\line(0,1){6}}
    \put(0,-0.1){\makebox(0,0)[t]{0}}
    \put(0,5){\line(1,0){22}}
    \put(22,5){\line(0,1){1}}
    \put(10,5){\line(0,1){1}}
    \put(2,-0.1){\makebox(0,0)[t]{1}}
    \put(2,0){\line(0,1){4.8}}
    \put(2,5.2){\line(0,1){0.8}}
    \put(2,4){\line(1,0){22}}
    \put(24,4){\line(0,1){2}}
    \put(12,4){\line(0,1){0.8}}
    \put(12,5.2){\line(0,1){0.8}}    
    \put(4,-0.1){\makebox(0,0)[t]{2}}
    \put(4,0){\line(0,1){3.8}}
    \put(4,4.2){\line(0,1){0.6}}
    \put(4,5.2){\line(0,1){0.8}}
    \put(4,3){\line(1,0){22}}
    \put(26,3){\line(0,1){3}}
    \put(14,3){\line(0,1){0.8}}
    \put(14,4.2){\line(0,1){0.6}}    
    \put(14,5.2){\line(0,1){0.8}}    
    \put(6,-0.1){\makebox(0,0)[t]{3}}
    \put(6,0){\line(0,1){2.8}}
    \put(6,3.2){\line(0,1){0.6}}    
    \put(6,4.2){\line(0,1){0.6}}    
    \put(6,5.2){\line(0,1){0.8}}    
    \put(6,2){\line(1,0){22}}
    \put(28,2){\line(0,1){4}}
    \put(16,2){\line(0,1){0.8}}
    \put(16,3.2){\line(0,1){0.6}}
    \put(16,4.2){\line(0,1){0.6}}
    \put(16,5.2){\line(0,1){0.8}}
    \put(8,-0.1){\makebox(0,0)[t]{4}}
    \put(8,0){\line(0,1){1.8}}
    \put(8,2.2){\line(0,1){0.6}}    
    \put(8,3.2){\line(0,1){0.6}}    
    \put(8,4.2){\line(0,1){0.6}}    
    \put(8,5.2){\line(0,1){0.8}}    
    \put(8,1){\line(1,0){22}}
    \put(30,1){\line(0,1){5}}
    \put(18,1){\line(0,1){0.8}}
    \put(18,2.2){\line(0,1){0.6}}
    \put(18,3.2){\line(0,1){0.6}}
    \put(18,4.2){\line(0,1){0.6}}
    \put(18,5.2){\line(0,1){0.8}}
  \end{picture}
  \end{center}
  \caption{Encoding the numbers $0,\ldots,4$ with three possibility of
    code each. The base for the code is the number $p=10$, after which
    there are four slots, one for each number. (At least) four slots
    are present after the numbers $p_5=15$, $p_6=21$, and
    $p_7=28$. For example, the number 2 can be encoded as 12, 17, or
    23.}
  \label{encode}
\end{figure}

\example
Let $n=5$ and assume that we want to encode each one of the numbers
$0,\ldots,4$ in three possible ways, from which we can choose at
random. We generate $p_4=10$, $p_5=15$, $p_6=21$, and
$p_7=28$. Between these numbers we can encode the numbers $0,\ldots,4$
as in figure~\ref{encode}.

\separate

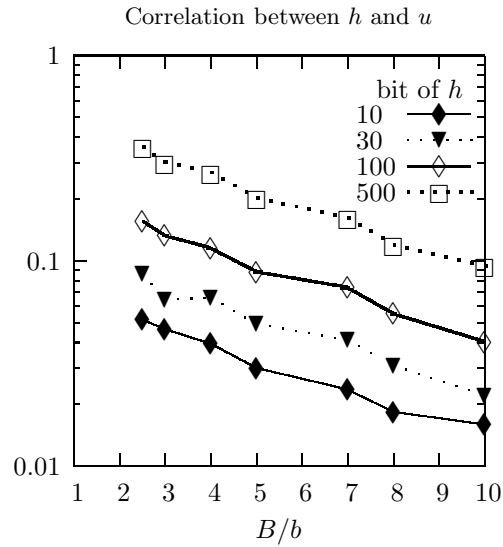
\begin{figure}
  \begin{center}
\setlength{\unitlength}{0.240900pt}
\ifx\plotpoint\undefined\newsavebox{\plotpoint}\fi
\sbox{\plotpoint}{\rule[-0.200pt]{0.400pt}{0.400pt}}%
\begin{picture}(900,900)(0,0)
\sbox{\plotpoint}{\rule[-0.200pt]{0.400pt}{0.400pt}}%
\put(172.0,131.0){\rule[-0.200pt]{4.818pt}{0.400pt}}
\put(152,131){\makebox(0,0)[r]{ 0.01}}
\put(797.0,131.0){\rule[-0.200pt]{4.818pt}{0.400pt}}
\put(172.0,228.0){\rule[-0.200pt]{2.409pt}{0.400pt}}
\put(807.0,228.0){\rule[-0.200pt]{2.409pt}{0.400pt}}
\put(172.0,285.0){\rule[-0.200pt]{2.409pt}{0.400pt}}
\put(807.0,285.0){\rule[-0.200pt]{2.409pt}{0.400pt}}
\put(172.0,325.0){\rule[-0.200pt]{2.409pt}{0.400pt}}
\put(807.0,325.0){\rule[-0.200pt]{2.409pt}{0.400pt}}
\put(172.0,356.0){\rule[-0.200pt]{2.409pt}{0.400pt}}
\put(807.0,356.0){\rule[-0.200pt]{2.409pt}{0.400pt}}
\put(172.0,382.0){\rule[-0.200pt]{2.409pt}{0.400pt}}
\put(807.0,382.0){\rule[-0.200pt]{2.409pt}{0.400pt}}
\put(172.0,404.0){\rule[-0.200pt]{2.409pt}{0.400pt}}
\put(807.0,404.0){\rule[-0.200pt]{2.409pt}{0.400pt}}
\put(172.0,422.0){\rule[-0.200pt]{2.409pt}{0.400pt}}
\put(807.0,422.0){\rule[-0.200pt]{2.409pt}{0.400pt}}
\put(172.0,439.0){\rule[-0.200pt]{2.409pt}{0.400pt}}
\put(807.0,439.0){\rule[-0.200pt]{2.409pt}{0.400pt}}
\put(172.0,453.0){\rule[-0.200pt]{4.818pt}{0.400pt}}
\put(152,453){\makebox(0,0)[r]{ 0.1}}
\put(797.0,453.0){\rule[-0.200pt]{4.818pt}{0.400pt}}
\put(172.0,551.0){\rule[-0.200pt]{2.409pt}{0.400pt}}
\put(807.0,551.0){\rule[-0.200pt]{2.409pt}{0.400pt}}
\put(172.0,607.0){\rule[-0.200pt]{2.409pt}{0.400pt}}
\put(807.0,607.0){\rule[-0.200pt]{2.409pt}{0.400pt}}
\put(172.0,648.0){\rule[-0.200pt]{2.409pt}{0.400pt}}
\put(807.0,648.0){\rule[-0.200pt]{2.409pt}{0.400pt}}
\put(172.0,679.0){\rule[-0.200pt]{2.409pt}{0.400pt}}
\put(807.0,679.0){\rule[-0.200pt]{2.409pt}{0.400pt}}
\put(172.0,704.0){\rule[-0.200pt]{2.409pt}{0.400pt}}
\put(807.0,704.0){\rule[-0.200pt]{2.409pt}{0.400pt}}
\put(172.0,726.0){\rule[-0.200pt]{2.409pt}{0.400pt}}
\put(807.0,726.0){\rule[-0.200pt]{2.409pt}{0.400pt}}
\put(172.0,745.0){\rule[-0.200pt]{2.409pt}{0.400pt}}
\put(807.0,745.0){\rule[-0.200pt]{2.409pt}{0.400pt}}
\put(172.0,761.0){\rule[-0.200pt]{2.409pt}{0.400pt}}
\put(807.0,761.0){\rule[-0.200pt]{2.409pt}{0.400pt}}
\put(172.0,776.0){\rule[-0.200pt]{4.818pt}{0.400pt}}
\put(152,776){\makebox(0,0)[r]{ 1}}
\put(797.0,776.0){\rule[-0.200pt]{4.818pt}{0.400pt}}
\put(172.0,131.0){\rule[-0.200pt]{0.400pt}{4.818pt}}
\put(172,90){\makebox(0,0){ 1}}
\put(172.0,756.0){\rule[-0.200pt]{0.400pt}{4.818pt}}
\put(244.0,131.0){\rule[-0.200pt]{0.400pt}{4.818pt}}
\put(244,90){\makebox(0,0){ 2}}
\put(244.0,756.0){\rule[-0.200pt]{0.400pt}{4.818pt}}
\put(315.0,131.0){\rule[-0.200pt]{0.400pt}{4.818pt}}
\put(315,90){\makebox(0,0){ 3}}
\put(315.0,756.0){\rule[-0.200pt]{0.400pt}{4.818pt}}
\put(387.0,131.0){\rule[-0.200pt]{0.400pt}{4.818pt}}
\put(387,90){\makebox(0,0){ 4}}
\put(387.0,756.0){\rule[-0.200pt]{0.400pt}{4.818pt}}
\put(459.0,131.0){\rule[-0.200pt]{0.400pt}{4.818pt}}
\put(459,90){\makebox(0,0){ 5}}
\put(459.0,756.0){\rule[-0.200pt]{0.400pt}{4.818pt}}
\put(530.0,131.0){\rule[-0.200pt]{0.400pt}{4.818pt}}
\put(530,90){\makebox(0,0){ 6}}
\put(530.0,756.0){\rule[-0.200pt]{0.400pt}{4.818pt}}
\put(602.0,131.0){\rule[-0.200pt]{0.400pt}{4.818pt}}
\put(602,90){\makebox(0,0){ 7}}
\put(602.0,756.0){\rule[-0.200pt]{0.400pt}{4.818pt}}
\put(674.0,131.0){\rule[-0.200pt]{0.400pt}{4.818pt}}
\put(674,90){\makebox(0,0){ 8}}
\put(674.0,756.0){\rule[-0.200pt]{0.400pt}{4.818pt}}
\put(745.0,131.0){\rule[-0.200pt]{0.400pt}{4.818pt}}
\put(745,90){\makebox(0,0){ 9}}
\put(745.0,756.0){\rule[-0.200pt]{0.400pt}{4.818pt}}
\put(817.0,131.0){\rule[-0.200pt]{0.400pt}{4.818pt}}
\put(817,90){\makebox(0,0){ 10}}
\put(817.0,756.0){\rule[-0.200pt]{0.400pt}{4.818pt}}
\put(172.0,131.0){\rule[-0.200pt]{0.400pt}{155.380pt}}
\put(172.0,131.0){\rule[-0.200pt]{155.380pt}{0.400pt}}
\put(817.0,131.0){\rule[-0.200pt]{0.400pt}{155.380pt}}
\put(172.0,776.0){\rule[-0.200pt]{155.380pt}{0.400pt}}
\put(494,29){\makebox(0,0){$B/b$}}
\put(494,838){\makebox(0,0){\small Correlation between $h$ and $u$}}
\put(716,724){\makebox(0,0){bit of $h$}}
\put(616,684){\makebox(0,0)[l]{10}}
\put(696.0,684.0){\rule[-0.200pt]{24.090pt}{0.400pt}}
\put(280,361){\usebox{\plotpoint}}
\multiput(280.00,359.92)(1.181,-0.494){27}{\rule{1.033pt}{0.119pt}}
\multiput(280.00,360.17)(32.855,-15.000){2}{\rule{0.517pt}{0.400pt}}
\multiput(315.00,344.92)(1.581,-0.496){43}{\rule{1.352pt}{0.120pt}}
\multiput(315.00,345.17)(69.193,-23.000){2}{\rule{0.676pt}{0.400pt}}
\multiput(387.00,321.92)(0.951,-0.498){73}{\rule{0.858pt}{0.120pt}}
\multiput(387.00,322.17)(70.219,-38.000){2}{\rule{0.429pt}{0.400pt}}
\multiput(459.00,283.92)(2.119,-0.498){65}{\rule{1.782pt}{0.120pt}}
\multiput(459.00,284.17)(139.301,-34.000){2}{\rule{0.891pt}{0.400pt}}
\multiput(602.00,249.92)(1.033,-0.498){67}{\rule{0.923pt}{0.120pt}}
\multiput(602.00,250.17)(70.085,-35.000){2}{\rule{0.461pt}{0.400pt}}
\multiput(674.00,214.92)(3.631,-0.496){37}{\rule{2.960pt}{0.119pt}}
\multiput(674.00,215.17)(136.856,-20.000){2}{\rule{1.480pt}{0.400pt}}
\put(280,361){\makebox(0,0){$\blacklozenge$}}
\put(315,346){\makebox(0,0){$\blacklozenge$}}
\put(387,323){\makebox(0,0){$\blacklozenge$}}
\put(459,285){\makebox(0,0){$\blacklozenge$}}
\put(602,251){\makebox(0,0){$\blacklozenge$}}
\put(674,216){\makebox(0,0){$\blacklozenge$}}
\put(817,196){\makebox(0,0){$\blacklozenge$}}
\put(746,684){\makebox(0,0){$\blacklozenge$}}
\put(616,643){\makebox(0,0)[l]{30}}
\multiput(696,643)(20.756,0.000){5}{\usebox{\plotpoint}}
\put(796,643){\usebox{\plotpoint}}
\put(280,433){\usebox{\plotpoint}}
\multiput(280,433)(13.668,-15.620){3}{\usebox{\plotpoint}}
\multiput(315,393)(20.747,0.576){4}{\usebox{\plotpoint}}
\multiput(387,395)(18.036,-10.271){4}{\usebox{\plotpoint}}
\multiput(459,354)(20.445,-3.574){7}{\usebox{\plotpoint}}
\multiput(602,329)(18.144,-10.080){3}{\usebox{\plotpoint}}
\multiput(674,289)(19.718,-6.481){8}{\usebox{\plotpoint}}
\put(817,242){\usebox{\plotpoint}}
\put(280,433){\makebox(0,0){$\blacktriangledown$}}
\put(315,393){\makebox(0,0){$\blacktriangledown$}}
\put(387,395){\makebox(0,0){$\blacktriangledown$}}
\put(459,354){\makebox(0,0){$\blacktriangledown$}}
\put(602,329){\makebox(0,0){$\blacktriangledown$}}
\put(674,289){\makebox(0,0){$\blacktriangledown$}}
\put(817,242){\makebox(0,0){$\blacktriangledown$}}
\put(746,643){\makebox(0,0){$\blacktriangledown$}}
\sbox{\plotpoint}{\rule[-0.400pt]{0.800pt}{0.800pt}}%
\sbox{\plotpoint}{\rule[-0.200pt]{0.400pt}{0.400pt}}%
\put(616,602){\makebox(0,0)[l]{100}}
\sbox{\plotpoint}{\rule[-0.400pt]{0.800pt}{0.800pt}}%
\put(696.0,602.0){\rule[-0.400pt]{24.090pt}{0.800pt}}
\put(280,516){\usebox{\plotpoint}}
\multiput(280.00,514.09)(0.766,-0.505){39}{\rule{1.417pt}{0.122pt}}
\multiput(280.00,514.34)(32.058,-23.000){2}{\rule{0.709pt}{0.800pt}}
\multiput(315.00,491.09)(1.850,-0.505){33}{\rule{3.080pt}{0.122pt}}
\multiput(315.00,491.34)(65.607,-20.000){2}{\rule{1.540pt}{0.800pt}}
\multiput(387.00,471.09)(0.980,-0.503){67}{\rule{1.757pt}{0.121pt}}
\multiput(387.00,471.34)(68.354,-37.000){2}{\rule{0.878pt}{0.800pt}}
\multiput(459.00,434.09)(2.933,-0.504){43}{\rule{4.776pt}{0.121pt}}
\multiput(459.00,434.34)(133.087,-25.000){2}{\rule{2.388pt}{0.800pt}}
\multiput(602.00,409.09)(0.883,-0.502){75}{\rule{1.605pt}{0.121pt}}
\multiput(602.00,409.34)(68.669,-41.000){2}{\rule{0.802pt}{0.800pt}}
\multiput(674.00,368.09)(1.642,-0.502){81}{\rule{2.800pt}{0.121pt}}
\multiput(674.00,368.34)(137.188,-44.000){2}{\rule{1.400pt}{0.800pt}}
\put(280,516){\makebox(0,0){$\lozenge$}}
\put(315,493){\makebox(0,0){$\lozenge$}}
\put(387,473){\makebox(0,0){$\lozenge$}}
\put(459,436){\makebox(0,0){$\lozenge$}}
\put(602,411){\makebox(0,0){$\lozenge$}}
\put(674,370){\makebox(0,0){$\lozenge$}}
\put(817,326){\makebox(0,0){$\lozenge$}}
\put(746,602){\makebox(0,0){$\lozenge$}}
\sbox{\plotpoint}{\rule[-0.500pt]{1.000pt}{1.000pt}}%
\sbox{\plotpoint}{\rule[-0.200pt]{0.400pt}{0.400pt}}%
\put(616,561){\makebox(0,0)[l]{500}}
\sbox{\plotpoint}{\rule[-0.500pt]{1.000pt}{1.000pt}}%
\multiput(696,561)(20.756,0.000){5}{\usebox{\plotpoint}}
\put(796,561){\usebox{\plotpoint}}
\put(280,632){\usebox{\plotpoint}}
\multiput(280,632)(17.118,-11.738){3}{\usebox{\plotpoint}}
\multiput(315,608)(20.261,-4.503){3}{\usebox{\plotpoint}}
\multiput(387,592)(18.144,-10.080){4}{\usebox{\plotpoint}}
\multiput(459,552)(20.284,-4.397){7}{\usebox{\plotpoint}}
\multiput(602,521)(17.819,-10.642){4}{\usebox{\plotpoint}}
\multiput(674,478)(20.224,-4.667){7}{\usebox{\plotpoint}}
\put(817,445){\usebox{\plotpoint}}
\put(280,632){\raisebox{-.8pt}{\makebox(0,0){$\Box$}}}
\put(315,608){\raisebox{-.8pt}{\makebox(0,0){$\Box$}}}
\put(387,592){\raisebox{-.8pt}{\makebox(0,0){$\Box$}}}
\put(459,552){\raisebox{-.8pt}{\makebox(0,0){$\Box$}}}
\put(602,521){\raisebox{-.8pt}{\makebox(0,0){$\Box$}}}
\put(674,478){\raisebox{-.8pt}{\makebox(0,0){$\Box$}}}
\put(817,445){\raisebox{-.8pt}{\makebox(0,0){$\Box$}}}
\put(746,561){\raisebox{-.8pt}{\makebox(0,0){$\Box$}}}
\sbox{\plotpoint}{\rule[-0.200pt]{0.400pt}{0.400pt}}%
\put(172.0,131.0){\rule[-0.200pt]{0.400pt}{155.380pt}}
\put(172.0,131.0){\rule[-0.200pt]{155.380pt}{0.400pt}}
\put(817.0,131.0){\rule[-0.200pt]{0.400pt}{155.380pt}}
\put(172.0,776.0){\rule[-0.200pt]{155.380pt}{0.400pt}}
\end{picture}
  \end{center}
  \caption{The correlation between $h$ and $u$ with $b$ and $B$ bits,
    respectively, for the quadratic randomization scheme (insertion of
    random bits in fixed position results in a correlation of $1$ in
    all cases). The curves are parametrized by the number of bits of
    $h$ and plotted as a function of $B/b$, where $B\ge{2b}$ is the
    number of bits of $u$.}
  \label{correl1}
\end{figure}

Suppose we want to encode the numbers $0,\ldots,M-1$ and we want at
least $M$ possible codes for each number. Then we need to consider our
sequence up to the point $p_{2M}$. Note that if $M$ is represented
using $b$ bits, then $p_{2M}$ is represented using $2b$ bits. The
encoding function is
\single
{\tt
\cb
encode(n,M) \\
1. \> k $\leftarrow$ rnd(n,2M-1); \\
2. \> cod $\leftarrow$ k(k+1)/2 + n; \\
3. \> \cmd{return} cod
\ce
}
\double
The inverse function identifies the number $p_n$ used in encoding and
returns the coded value as the difference between the code and $p_n$:
\single
{\tt
\cb
decode(u) \\
1. \> k $\leftarrow$ max$\Bigl\{$ n | n(n+1)/2 $\le$ u $\Bigr\}$ \\
2. \> \cmd{return} u - k(k+1)/2;
\ce
}
\double
This function breaks the correlation between $h$ and
$u$=encode($h$,M).if $h$ has $b$ bits, and $u$ has $B>b$ bits, then
Figure~\ref{correl1} shows the correlation found by CCA as a function
of $B/b$.
Note that in these curves the correlation increases as $B/b$
increases, a somewhat counterintuitive behavior. Analysis of the
behavior allowed us to determine that the correlation was catching on
to the fact that some of the high bits of the code were often zero,
and correlated this with the zeros in $h$. That is, the increased
correlation is due to systematic confusion, and not to a better
detection of $h$.

\section{Cipher and Key distribution}
The value $u$ must now be encoded using a suitable cipher so that the
result can be published. Thanks to the function \texttt{encode}, for
each message $h$ that we want to publish $T$ times associated to the
same manifest content $m$, we now have $T$ representations
$[u_1,\ldots,u_T]$ de-correlated from $h$ and with low mutual
information.

We now use an encryption scheme which depends on
a key $k$ so that the encrypted values $c(u_i,k)$ can be published
together with the manifest content. The characteristics of our problem
prevent us from using a public key method such as RSA
\cite{rivest:78}. In public key methods, the owner of the key
distributes freely an encoding key $k_1$, but keeps secret the key
$k_2$ necessary for decoding. That is, given the message $u$, anybody
can create the cipher $c(u,k_1)$, but only the owner of the decoding key can
invert the cipher and reconstruct $u=c'(c(u,k_1),k_2)$. We are in a
different situation: we have a group of trusted people
$\{t_1,\ldots,t_n\}$ whom we want to be able to decode the message, so
that the decoding key must be distributed.

We use a symmetric encryption method such as one-time pad
\cite{schneier:07}; given a message $u$ of $B$ bits,
$u=u_0\cdots{u}_{B-1}$, $u_i\in\{0,1\}$, we ue a $B$-bit key
$k=k_0\cdots{k}_{B-1}$ and create the cipher $c=c_0\cdots{c}_{B-1}$ as
the bit-wise exclusive or between $u$ and $k$:
\begin{equation}
  c_i = u_i \oplus k_i
\end{equation}
If the receiver has the key $k$, the message $u$ can be decoded with
another bit-wise exclusive or. The problem is now reduced to the safe
distribution of $k$ to the trusted group on an insecure connection.

For this, we use a generalization of the Diffie-Hellman protocol
\cite{maurer:00} to distribute a key among $n$ participants,
originating from a central distributor (the person who will publish
the tags). Table~\ref{stuff} resumes the elements that we use in our
protocol.
\begin{table}
  \begin{center}
    \begin{tabular}{||c|p{20em}||}
      \hline
      \hline
      $G$  & Algebraic group (finite, cyclic). \\
      $q$  & Order of $G$. \\
      $n$  & Number of participants who will receive the key. \\
      $M_i$ & $i$th participant, $i\in\{0,\ldots,n-1\}$. \\
      $A$   & Distributor of the key. \\
      $\alpha$ & Generator of $G$. \\
      $N_i$ & Private key of $M_i$. \\
      $N_a$ & Private key of $A$. \\
      $K$   & Shared key. \\
      \hline
      \hline
    \end{tabular}
  \end{center}
  \caption{The quantities used in the key distribution protocol.}
  \label{stuff}
\end{table}
Additionally, as a matter of notation, we shall indicate as
$N_1\cdots\bar{N_i}\cdots{N_k}$ the product
$N_1\cdot\cdots{N}_{i-1}N_{i+1}\cdots{N_k}$, that is, the
multiplication of a number of terms from which $N_i$ is excluded. Also,
to avoid the use of large superscripts, we shall indicate the powers
$\alpha^N$ as $\alpha\!\uparrow\!{N}$. All the powers are done on the
cyclic group, that is, assuming that the group is isomorphic to
${\mathbb{Z}}_q$, all the multiplications are modulo $q$.

The protocol consists of five steps:
\begin{description}
\item[i)] $A$ collects the value $\alpha\!\uparrow\!N_1\cdots{N_n}$;
  to this end:
  \begin{description}
    \item[a)] $M_1$ computes $\alpha\!\uparrow\!{N_1}$ and sends it to
      $A$
    \item[b)] for $i=2,\ldots,n$
      \begin{description}
        \item[b.1)] $A$ sends $\alpha\!\uparrow\!N_1\cdots{N}_{i-1}$ to $M_i$
        \item[b.2)] $M_i$ elevates the value received to the $N_i$th
          power, thus computing $\alpha\!\uparrow\!N_1\cdots{N_i}$, and
          sends it to $A$.
      \end{description}
  \end{description}
\item[ii)] $A$ broadcasts $\alpha\!\uparrow\!N_1\cdots{N_n}$ to all the
  members $M_i$
\item[iii)] Each $M_i$ computes $N_i^{-1}$ and elevates
  $\alpha\!\uparrow\!N_1\cdots{N_n}$ to this power, obtaining
  $\alpha\!\uparrow\!N_1\cdots\bar{N_i}\cdots{N_n}$, and sends the
  result to $A$.
\item[iv)] $A$ elevates all these values to the private key $N_a$, and
  sends, to each $M_i$, the value
  $\alpha\!\uparrow\!N_1\cdots\bar{N_i}\cdots{N_n}N_a$
\item[v)] Each $M_i$ elevates the value received to $N_i$, obtaining
  the shared key  $K=\alpha\!\uparrow\!N_1\cdots{N_n}N_a$.
\end{description}

At this point all the members $M_i$, as well as the producer $A$ share
the same key $K$. The protocol requires, in total, the transmission of
$5n$ partial keys (numbers of $b$ bits each), so its complexity is
$O(n)$.

\subsection{Security of the protocol}
The security of our key interchange protocol is based on the
Computational Diffie-Hellman Hypothesis (CDH, \cite{maurer:94}).

Consider a group $G$ of order $q$ (we assume $q$ to be prime), a
generator $g\in{G}$, and an element $h\in{G}$. The discrete logarithm
problem consists in finding $\lambda\dfeq\log_g{h}$ such that
$g^\lambda=h$. An algorithm solves the discrete logarithm problem if,
for all $g\ne{1}$ and $h$, determines $\log_g{h}$ with a success
probability $p>p_0$, for $p_0$ given. The \emph{Discrete Logarithm
  hypothesis} is that no algorithm polynomial in the number of bits of
$h$ can solve the discrete logarithm problem \cite{maurer:94}.

The Diffie-Hellman problem with base $g$ consists in, given
$g_1=g^\alpha$ and $g_2=g^\beta$, in computing $g^{\alpha\beta}$. The
CDH hypothesis states that no algorithm with running time polynomial
in the number of bits of $g,\alpha,\beta$ can solve this problem with
a probability of success significantly greater than chance.

The \emph{CDH hypothesis in decision form} (DDH, \cite{boneh:98})
states that the tuples $(g^\alpha,g^\beta,g^{\alpha\beta})$ and
$(g^\alpha,g^\beta,g^\chi)$, where $\alpha,\beta,\chi$ are independent
and randomly chosen are computationally indistinguishable, that is,
given any decision algorithm $A$ that stops on $T$ or $F$, it is
\begin{equation}
  \Bigl| {\mathbb{P}}\{A(g^\alpha,g^\beta,g^{\alpha\beta})=T\} - {\mathbb{P}}\{A(g^\alpha,g^\beta,g^{\chi})=T\} \Bigl| \sim 0
\end{equation}

We indicate with $A\sim{B}$ the fact that the sets $A$ and $B$ are
computationally indistinguishable. Also, we define the following
quantities; for $X=\{N_1,\ldots,N_n\}$
\begin{equation}
  V(m,X) = \Bigl\{ \alpha\!\uparrow{N}_{i_1}\cdots{N}_{i_s} \Bigl| \{i_1,\ldots,i_s\} \subset \{1,\ldots,m\} \Bigr\}
\end{equation}
(note that subsethood is strict: $\alpha\!\uparrow\!N_1\cdots{N_m}\not\in{V}(m,X)$).
\begin{equation}
  \begin{aligned}
    K(X) &= \{\alpha\!\uparrow\!N_1\cdots{N_m}\} \\
    A_m  &= V(m,X) \cup \{y\} \ \ \mbox{$y$ random} \\
    D_m  &= V(m,X) \cup K(X)
  \end{aligned}
\end{equation}
$V(m,X)$ contains all the information that is exchanged during the
protocol. If $D_m$ (the information plus the key) is computationally
indistinguishable from $A_m$ (the information plus a random element),
the protocol is safe. The safety is then established by the following
theorem

\begin{theorem}
  If DDH holds (viz.\ if $A_2\sim{D_2}$) then $A_m\sim{D_m}$ for all $m$.
\end{theorem}

\begin{proof}
  The case $m=2$ follows directly from DDH, so assume that
  $A_{m-1}\sim{D}_{m-1}$.

  Let $X=\{N_3,\ldots,N_m\}$, then
  \begin{equation}
    \begin{aligned}
      V(m,\{N_1,N_2,X\}) = &   
      V(m-1,\{N_1,X\}) \cup 
      K(\{N_1,X\}) \\
      & \cup  
      V(m-1,\{N_2,X\}) \cup 
      K(\{N_2,X\}) \\
      & \cup 
      V(m-1,\{N_1,N_2,X\})
    \end{aligned}
  \end{equation}
  Expanding $A_m$ and $D_m$ in a similar way, we have
  \begin{equation}
    \begin{aligned}
      A_m =& 
      V(m-1,\{N_1,X\}) \cup K(\{N_1,X\}) \\\cup  V(m-1,\{N_2,X\}) \cup K(\{N_2,X\}) \\
      & \hspace{3em}\cup 
      V(m-1,\{N_1,N_2,X\}) \cup \{y\} \\
      D_m =& 
      V(m-1,\{N_1,X\}) \cup K(\{N_1,X\}) \cup  V(m-1,\{N_2,X\}) \cup K(\{N_2,X\}) \\
      & \hspace{3em}\cup 
      V(m-1,\{N_1,N_2,X\}) \cup K(\{N_1,N_2,X\}) 
    \end{aligned}
  \end{equation}
  Define now $B_m$ and $C_m$ as follows:
  \begin{equation}
    \begin{aligned}
      B_m =& 
      V(m-1,\{N_1,X\}) \cup K(\{N_1,X\}) \\
      & \hspace{3em}\cup  V(m-1,\{N_2,X\}) \cup K(\{N_2,X\}) \\
      & \hspace{3em}\cup V(m-1,\{c,X\}) \cup \{y\} \\
      C_m =& 
      V(m-1,\{N_1,X\}) \cup K(\{N_1,X\}) \\
      & \hspace{3em}\cup  V(m-1,\{N_2,X\}) \cup K(\{N_2,X\}) \\
      & \hspace{3em}\cup 
      V(m-1,\{c,X\}) \cup K(\{c,X\}) 
    \end{aligned}
  \end{equation}
  where $c$ is a random element.

  We show first that $A_m\sim{B}_m$. Suppose that an algorithm
  distinguishes between them. Since
  $\alpha\!\uparrow\!{N_1N_2}\in{V}(m-1,\{N_1,N_2,X\})\subset{A_m}$
  and $\alpha\!\uparrow\!c\in{V}(m-1,\{c,X\})\subset{B_m}$, this would
  allow the program to distinguish between $A_2$ and $B_2$,
  contradicting DDH.

  Consider now $B_m$ and $C_m$. We have $y\in{B_m}$ and
  $K(\{c,X\})\in{C_m}$, and this is the only difference between the
  two so, if we can distinguish between $B_m$ and $C_m$ we can decide
  whether $y=K(\{c,X\})$. This is an instance of
  $A_{m-1}\sim{D_{m-1}}$ so distinguishability of $B_m$ and $C_m$
  contradicts the inductive hypothesis. Therefore $B_m\sim{C}_m$.

  Finally, consider $C_m$ and $D_m$. If one distinguishes the two,
  then one can distinguish between
  $\alpha\!\uparrow\!N_1N_2\in{V}(m-1,\{N_1,N_2,X\})\subset{D_m}$ and
  $\alpha\!\uparrow\!c\in{V}(m-1,\{c,X\})\subset{C_m}$, that is, we can
  distinguish between $C_2$ and $D_2$, violating DDH. Therefore
  $C_m\sim{D}_m$.

  By transitivity then $A_m\sim{D_m}$.
\end{proof}

\section{Remarks}
Our purpose is to prevent statistical attacks, that is, to prevent an
intruder from identifying that different copies of the same hidden
message $h$ are indeed the same message. In order to do this, our
strategy has been to reduce the correlation between $h$ and $c(u)$, by
encoding a randomized function of $h$: $u=u(h,r)$. Encryption will
help us in this: given any randomization $u=f(h,r)$ ($r$ random),
encryption will in general reduce the correlation of $u$ with $h$,
that is, in general, $\rho(c(u),h)<\rho(u,h)$.

\begin{figure*}[htbp]
  \begin{center}
    \begin{tabular}{cccl}

      \vspace{-2.5em}

      \hspace{-6em}\begin{rotate}{90}\hspace{4.5em}{\small  length: b=10 bits}\end{rotate}& \hspace{-3em}
\setlength{\unitlength}{0.240900pt}
\ifx\plotpoint\undefined\newsavebox{\plotpoint}\fi
\sbox{\plotpoint}{\rule[-0.200pt]{0.400pt}{0.400pt}}%
\begin{picture}(675,600)(0,0)
\sbox{\plotpoint}{\rule[-0.200pt]{0.400pt}{0.400pt}}%
\put(150.0,82.0){\rule[-0.200pt]{4.818pt}{0.400pt}}
\put(130,82){\makebox(0,0)[r]{ 0.01}}
\put(594.0,82.0){\rule[-0.200pt]{4.818pt}{0.400pt}}
\put(150.0,130.0){\rule[-0.200pt]{2.409pt}{0.400pt}}
\put(604.0,130.0){\rule[-0.200pt]{2.409pt}{0.400pt}}
\put(150.0,158.0){\rule[-0.200pt]{2.409pt}{0.400pt}}
\put(604.0,158.0){\rule[-0.200pt]{2.409pt}{0.400pt}}
\put(150.0,178.0){\rule[-0.200pt]{2.409pt}{0.400pt}}
\put(604.0,178.0){\rule[-0.200pt]{2.409pt}{0.400pt}}
\put(150.0,193.0){\rule[-0.200pt]{2.409pt}{0.400pt}}
\put(604.0,193.0){\rule[-0.200pt]{2.409pt}{0.400pt}}
\put(150.0,206.0){\rule[-0.200pt]{2.409pt}{0.400pt}}
\put(604.0,206.0){\rule[-0.200pt]{2.409pt}{0.400pt}}
\put(150.0,216.0){\rule[-0.200pt]{2.409pt}{0.400pt}}
\put(604.0,216.0){\rule[-0.200pt]{2.409pt}{0.400pt}}
\put(150.0,226.0){\rule[-0.200pt]{2.409pt}{0.400pt}}
\put(604.0,226.0){\rule[-0.200pt]{2.409pt}{0.400pt}}
\put(150.0,234.0){\rule[-0.200pt]{2.409pt}{0.400pt}}
\put(604.0,234.0){\rule[-0.200pt]{2.409pt}{0.400pt}}
\put(150.0,241.0){\rule[-0.200pt]{4.818pt}{0.400pt}}
\put(130,241){\makebox(0,0)[r]{ 0.1}}
\put(594.0,241.0){\rule[-0.200pt]{4.818pt}{0.400pt}}
\put(150.0,289.0){\rule[-0.200pt]{2.409pt}{0.400pt}}
\put(604.0,289.0){\rule[-0.200pt]{2.409pt}{0.400pt}}
\put(150.0,317.0){\rule[-0.200pt]{2.409pt}{0.400pt}}
\put(604.0,317.0){\rule[-0.200pt]{2.409pt}{0.400pt}}
\put(150.0,337.0){\rule[-0.200pt]{2.409pt}{0.400pt}}
\put(604.0,337.0){\rule[-0.200pt]{2.409pt}{0.400pt}}
\put(150.0,352.0){\rule[-0.200pt]{2.409pt}{0.400pt}}
\put(604.0,352.0){\rule[-0.200pt]{2.409pt}{0.400pt}}
\put(150.0,365.0){\rule[-0.200pt]{2.409pt}{0.400pt}}
\put(604.0,365.0){\rule[-0.200pt]{2.409pt}{0.400pt}}
\put(150.0,375.0){\rule[-0.200pt]{2.409pt}{0.400pt}}
\put(604.0,375.0){\rule[-0.200pt]{2.409pt}{0.400pt}}
\put(150.0,385.0){\rule[-0.200pt]{2.409pt}{0.400pt}}
\put(604.0,385.0){\rule[-0.200pt]{2.409pt}{0.400pt}}
\put(150.0,393.0){\rule[-0.200pt]{2.409pt}{0.400pt}}
\put(604.0,393.0){\rule[-0.200pt]{2.409pt}{0.400pt}}
\put(150.0,400.0){\rule[-0.200pt]{4.818pt}{0.400pt}}
\put(130,400){\makebox(0,0)[r]{ 1}}
\put(594.0,400.0){\rule[-0.200pt]{4.818pt}{0.400pt}}
\put(150.0,448.0){\rule[-0.200pt]{2.409pt}{0.400pt}}
\put(604.0,448.0){\rule[-0.200pt]{2.409pt}{0.400pt}}
\put(150.0,476.0){\rule[-0.200pt]{2.409pt}{0.400pt}}
\put(604.0,476.0){\rule[-0.200pt]{2.409pt}{0.400pt}}
\put(150.0,496.0){\rule[-0.200pt]{2.409pt}{0.400pt}}
\put(604.0,496.0){\rule[-0.200pt]{2.409pt}{0.400pt}}
\put(150.0,511.0){\rule[-0.200pt]{2.409pt}{0.400pt}}
\put(604.0,511.0){\rule[-0.200pt]{2.409pt}{0.400pt}}
\put(150.0,524.0){\rule[-0.200pt]{2.409pt}{0.400pt}}
\put(604.0,524.0){\rule[-0.200pt]{2.409pt}{0.400pt}}
\put(150.0,534.0){\rule[-0.200pt]{2.409pt}{0.400pt}}
\put(604.0,534.0){\rule[-0.200pt]{2.409pt}{0.400pt}}
\put(150.0,544.0){\rule[-0.200pt]{2.409pt}{0.400pt}}
\put(604.0,544.0){\rule[-0.200pt]{2.409pt}{0.400pt}}
\put(150.0,552.0){\rule[-0.200pt]{2.409pt}{0.400pt}}
\put(604.0,552.0){\rule[-0.200pt]{2.409pt}{0.400pt}}
\put(150.0,559.0){\rule[-0.200pt]{4.818pt}{0.400pt}}
\put(130,559){\makebox(0,0)[r]{ 10}}
\put(594.0,559.0){\rule[-0.200pt]{4.818pt}{0.400pt}}
\put(150.0,82.0){\rule[-0.200pt]{0.400pt}{4.818pt}}
\put(150,41){\makebox(0,0){ }}
\put(150.0,539.0){\rule[-0.200pt]{0.400pt}{4.818pt}}
\put(202.0,82.0){\rule[-0.200pt]{0.400pt}{4.818pt}}
\put(202,41){\makebox(0,0){ }}
\put(202.0,539.0){\rule[-0.200pt]{0.400pt}{4.818pt}}
\put(253.0,82.0){\rule[-0.200pt]{0.400pt}{4.818pt}}
\put(253,41){\makebox(0,0){ }}
\put(253.0,539.0){\rule[-0.200pt]{0.400pt}{4.818pt}}
\put(305.0,82.0){\rule[-0.200pt]{0.400pt}{4.818pt}}
\put(305,41){\makebox(0,0){ }}
\put(305.0,539.0){\rule[-0.200pt]{0.400pt}{4.818pt}}
\put(356.0,82.0){\rule[-0.200pt]{0.400pt}{4.818pt}}
\put(356,41){\makebox(0,0){ }}
\put(356.0,539.0){\rule[-0.200pt]{0.400pt}{4.818pt}}
\put(408.0,82.0){\rule[-0.200pt]{0.400pt}{4.818pt}}
\put(408,41){\makebox(0,0){ }}
\put(408.0,539.0){\rule[-0.200pt]{0.400pt}{4.818pt}}
\put(459.0,82.0){\rule[-0.200pt]{0.400pt}{4.818pt}}
\put(459,41){\makebox(0,0){ }}
\put(459.0,539.0){\rule[-0.200pt]{0.400pt}{4.818pt}}
\put(511.0,82.0){\rule[-0.200pt]{0.400pt}{4.818pt}}
\put(511,41){\makebox(0,0){ }}
\put(511.0,539.0){\rule[-0.200pt]{0.400pt}{4.818pt}}
\put(562.0,82.0){\rule[-0.200pt]{0.400pt}{4.818pt}}
\put(562,41){\makebox(0,0){ }}
\put(562.0,539.0){\rule[-0.200pt]{0.400pt}{4.818pt}}
\put(614.0,82.0){\rule[-0.200pt]{0.400pt}{4.818pt}}
\put(614,41){\makebox(0,0){ }}
\put(614.0,539.0){\rule[-0.200pt]{0.400pt}{4.818pt}}
\put(150.0,82.0){\rule[-0.200pt]{0.400pt}{114.909pt}}
\put(150.0,82.0){\rule[-0.200pt]{111.778pt}{0.400pt}}
\put(614.0,82.0){\rule[-0.200pt]{0.400pt}{114.909pt}}
\put(150.0,559.0){\rule[-0.200pt]{111.778pt}{0.400pt}}
\put(227,147){\usebox{\plotpoint}}
\multiput(253.00,145.92)(1.313,-0.496){37}{\rule{1.140pt}{0.119pt}}
\multiput(253.00,146.17)(49.634,-20.000){2}{\rule{0.570pt}{0.400pt}}
\multiput(305.00,125.92)(0.593,-0.498){83}{\rule{0.574pt}{0.120pt}}
\multiput(305.00,126.17)(49.808,-43.000){2}{\rule{0.287pt}{0.400pt}}
\multiput(356.00,84.58)(4.417,0.492){21}{\rule{3.533pt}{0.119pt}}
\multiput(356.00,83.17)(95.666,12.000){2}{\rule{1.767pt}{0.400pt}}
\multiput(459.00,94.92)(1.084,-0.494){25}{\rule{0.957pt}{0.119pt}}
\multiput(459.00,95.17)(28.013,-14.000){2}{\rule{0.479pt}{0.400pt}}
\put(227,147){\makebox(0,0){$\blacklozenge$}}
\put(253,147){\makebox(0,0){$\blacklozenge$}}
\put(305,127){\makebox(0,0){$\blacklozenge$}}
\put(356,84){\makebox(0,0){$\blacklozenge$}}
\put(459,96){\makebox(0,0){$\blacklozenge$}}
\put(227.0,147.0){\rule[-0.200pt]{6.263pt}{0.400pt}}
\put(227,159){\usebox{\plotpoint}}
\multiput(227,159)(18.275,9.840){2}{\usebox{\plotpoint}}
\multiput(253,173)(17.677,-10.878){3}{\usebox{\plotpoint}}
\multiput(305,141)(20.112,-5.127){2}{\usebox{\plotpoint}}
\multiput(356,128)(20.029,-5.445){6}{\usebox{\plotpoint}}
\multiput(459,100)(18.706,8.993){2}{\usebox{\plotpoint}}
\multiput(511,125)(19.766,-6.333){6}{\usebox{\plotpoint}}
\put(614,92){\usebox{\plotpoint}}
\put(227,159){\makebox(0,0){$\lozenge$}}
\put(253,173){\makebox(0,0){$\lozenge$}}
\put(305,141){\makebox(0,0){$\lozenge$}}
\put(356,128){\makebox(0,0){$\lozenge$}}
\put(459,100){\makebox(0,0){$\lozenge$}}
\put(511,125){\makebox(0,0){$\lozenge$}}
\put(614,92){\makebox(0,0){$\lozenge$}}
\sbox{\plotpoint}{\rule[-0.400pt]{0.800pt}{0.800pt}}%
\put(227,412){\usebox{\plotpoint}}
\multiput(227.00,413.41)(0.950,0.509){21}{\rule{1.686pt}{0.123pt}}
\multiput(227.00,410.34)(22.501,14.000){2}{\rule{0.843pt}{0.800pt}}
\multiput(253.00,424.08)(2.299,-0.511){17}{\rule{3.667pt}{0.123pt}}
\multiput(253.00,424.34)(44.390,-12.000){2}{\rule{1.833pt}{0.800pt}}
\multiput(305.00,415.41)(0.856,0.503){53}{\rule{1.560pt}{0.121pt}}
\multiput(305.00,412.34)(47.762,30.000){2}{\rule{0.780pt}{0.800pt}}
\multiput(356.00,442.09)(1.268,-0.502){75}{\rule{2.210pt}{0.121pt}}
\multiput(356.00,442.34)(98.414,-41.000){2}{\rule{1.105pt}{0.800pt}}
\multiput(459.00,404.41)(0.519,0.502){93}{\rule{1.032pt}{0.121pt}}
\multiput(459.00,401.34)(49.858,50.000){2}{\rule{0.516pt}{0.800pt}}
\multiput(511.00,454.41)(1.634,0.503){57}{\rule{2.775pt}{0.121pt}}
\multiput(511.00,451.34)(97.240,32.000){2}{\rule{1.388pt}{0.800pt}}
\put(227,412){\makebox(0,0){$\blacktriangledown$}}
\put(253,426){\makebox(0,0){$\blacktriangledown$}}
\put(305,414){\makebox(0,0){$\blacktriangledown$}}
\put(356,444){\makebox(0,0){$\blacktriangledown$}}
\put(459,403){\makebox(0,0){$\blacktriangledown$}}
\put(511,453){\makebox(0,0){$\blacktriangledown$}}
\put(614,485){\makebox(0,0){$\blacktriangledown$}}
\sbox{\plotpoint}{\rule[-0.200pt]{0.400pt}{0.400pt}}%
\put(150.0,82.0){\rule[-0.200pt]{0.400pt}{114.909pt}}
\put(150.0,82.0){\rule[-0.200pt]{111.778pt}{0.400pt}}
\put(614.0,82.0){\rule[-0.200pt]{0.400pt}{114.909pt}}
\put(150.0,559.0){\rule[-0.200pt]{111.778pt}{0.400pt}}
\end{picture} & %
\hspace{-2em}
\setlength{\unitlength}{0.240900pt}
\ifx\plotpoint\undefined\newsavebox{\plotpoint}\fi
\begin{picture}(600,600)(0,0)
\sbox{\plotpoint}{\rule[-0.200pt]{0.400pt}{0.400pt}}%
\put(70.0,82.0){\rule[-0.200pt]{4.818pt}{0.400pt}}
\put(50,82){\makebox(0,0)[r]{ }}
\put(519.0,82.0){\rule[-0.200pt]{4.818pt}{0.400pt}}
\put(70.0,130.0){\rule[-0.200pt]{2.409pt}{0.400pt}}
\put(529.0,130.0){\rule[-0.200pt]{2.409pt}{0.400pt}}
\put(70.0,158.0){\rule[-0.200pt]{2.409pt}{0.400pt}}
\put(529.0,158.0){\rule[-0.200pt]{2.409pt}{0.400pt}}
\put(70.0,178.0){\rule[-0.200pt]{2.409pt}{0.400pt}}
\put(529.0,178.0){\rule[-0.200pt]{2.409pt}{0.400pt}}
\put(70.0,193.0){\rule[-0.200pt]{2.409pt}{0.400pt}}
\put(529.0,193.0){\rule[-0.200pt]{2.409pt}{0.400pt}}
\put(70.0,206.0){\rule[-0.200pt]{2.409pt}{0.400pt}}
\put(529.0,206.0){\rule[-0.200pt]{2.409pt}{0.400pt}}
\put(70.0,216.0){\rule[-0.200pt]{2.409pt}{0.400pt}}
\put(529.0,216.0){\rule[-0.200pt]{2.409pt}{0.400pt}}
\put(70.0,226.0){\rule[-0.200pt]{2.409pt}{0.400pt}}
\put(529.0,226.0){\rule[-0.200pt]{2.409pt}{0.400pt}}
\put(70.0,234.0){\rule[-0.200pt]{2.409pt}{0.400pt}}
\put(529.0,234.0){\rule[-0.200pt]{2.409pt}{0.400pt}}
\put(70.0,241.0){\rule[-0.200pt]{4.818pt}{0.400pt}}
\put(50,241){\makebox(0,0)[r]{ }}
\put(519.0,241.0){\rule[-0.200pt]{4.818pt}{0.400pt}}
\put(70.0,289.0){\rule[-0.200pt]{2.409pt}{0.400pt}}
\put(529.0,289.0){\rule[-0.200pt]{2.409pt}{0.400pt}}
\put(70.0,317.0){\rule[-0.200pt]{2.409pt}{0.400pt}}
\put(529.0,317.0){\rule[-0.200pt]{2.409pt}{0.400pt}}
\put(70.0,337.0){\rule[-0.200pt]{2.409pt}{0.400pt}}
\put(529.0,337.0){\rule[-0.200pt]{2.409pt}{0.400pt}}
\put(70.0,352.0){\rule[-0.200pt]{2.409pt}{0.400pt}}
\put(529.0,352.0){\rule[-0.200pt]{2.409pt}{0.400pt}}
\put(70.0,365.0){\rule[-0.200pt]{2.409pt}{0.400pt}}
\put(529.0,365.0){\rule[-0.200pt]{2.409pt}{0.400pt}}
\put(70.0,375.0){\rule[-0.200pt]{2.409pt}{0.400pt}}
\put(529.0,375.0){\rule[-0.200pt]{2.409pt}{0.400pt}}
\put(70.0,385.0){\rule[-0.200pt]{2.409pt}{0.400pt}}
\put(529.0,385.0){\rule[-0.200pt]{2.409pt}{0.400pt}}
\put(70.0,393.0){\rule[-0.200pt]{2.409pt}{0.400pt}}
\put(529.0,393.0){\rule[-0.200pt]{2.409pt}{0.400pt}}
\put(70.0,400.0){\rule[-0.200pt]{4.818pt}{0.400pt}}
\put(50,400){\makebox(0,0)[r]{ }}
\put(519.0,400.0){\rule[-0.200pt]{4.818pt}{0.400pt}}
\put(70.0,448.0){\rule[-0.200pt]{2.409pt}{0.400pt}}
\put(529.0,448.0){\rule[-0.200pt]{2.409pt}{0.400pt}}
\put(70.0,476.0){\rule[-0.200pt]{2.409pt}{0.400pt}}
\put(529.0,476.0){\rule[-0.200pt]{2.409pt}{0.400pt}}
\put(70.0,496.0){\rule[-0.200pt]{2.409pt}{0.400pt}}
\put(529.0,496.0){\rule[-0.200pt]{2.409pt}{0.400pt}}
\put(70.0,511.0){\rule[-0.200pt]{2.409pt}{0.400pt}}
\put(529.0,511.0){\rule[-0.200pt]{2.409pt}{0.400pt}}
\put(70.0,524.0){\rule[-0.200pt]{2.409pt}{0.400pt}}
\put(529.0,524.0){\rule[-0.200pt]{2.409pt}{0.400pt}}
\put(70.0,534.0){\rule[-0.200pt]{2.409pt}{0.400pt}}
\put(529.0,534.0){\rule[-0.200pt]{2.409pt}{0.400pt}}
\put(70.0,544.0){\rule[-0.200pt]{2.409pt}{0.400pt}}
\put(529.0,544.0){\rule[-0.200pt]{2.409pt}{0.400pt}}
\put(70.0,552.0){\rule[-0.200pt]{2.409pt}{0.400pt}}
\put(529.0,552.0){\rule[-0.200pt]{2.409pt}{0.400pt}}
\put(70.0,559.0){\rule[-0.200pt]{4.818pt}{0.400pt}}
\put(50,559){\makebox(0,0)[r]{ }}
\put(519.0,559.0){\rule[-0.200pt]{4.818pt}{0.400pt}}
\put(70.0,82.0){\rule[-0.200pt]{0.400pt}{4.818pt}}
\put(70,41){\makebox(0,0){ }}
\put(70.0,539.0){\rule[-0.200pt]{0.400pt}{4.818pt}}
\put(122.0,82.0){\rule[-0.200pt]{0.400pt}{4.818pt}}
\put(122,41){\makebox(0,0){ }}
\put(122.0,539.0){\rule[-0.200pt]{0.400pt}{4.818pt}}
\put(174.0,82.0){\rule[-0.200pt]{0.400pt}{4.818pt}}
\put(174,41){\makebox(0,0){ }}
\put(174.0,539.0){\rule[-0.200pt]{0.400pt}{4.818pt}}
\put(226.0,82.0){\rule[-0.200pt]{0.400pt}{4.818pt}}
\put(226,41){\makebox(0,0){ }}
\put(226.0,539.0){\rule[-0.200pt]{0.400pt}{4.818pt}}
\put(278.0,82.0){\rule[-0.200pt]{0.400pt}{4.818pt}}
\put(278,41){\makebox(0,0){ }}
\put(278.0,539.0){\rule[-0.200pt]{0.400pt}{4.818pt}}
\put(331.0,82.0){\rule[-0.200pt]{0.400pt}{4.818pt}}
\put(331,41){\makebox(0,0){ }}
\put(331.0,539.0){\rule[-0.200pt]{0.400pt}{4.818pt}}
\put(383.0,82.0){\rule[-0.200pt]{0.400pt}{4.818pt}}
\put(383,41){\makebox(0,0){ }}
\put(383.0,539.0){\rule[-0.200pt]{0.400pt}{4.818pt}}
\put(435.0,82.0){\rule[-0.200pt]{0.400pt}{4.818pt}}
\put(435,41){\makebox(0,0){ }}
\put(435.0,539.0){\rule[-0.200pt]{0.400pt}{4.818pt}}
\put(487.0,82.0){\rule[-0.200pt]{0.400pt}{4.818pt}}
\put(487,41){\makebox(0,0){ }}
\put(487.0,539.0){\rule[-0.200pt]{0.400pt}{4.818pt}}
\put(539.0,82.0){\rule[-0.200pt]{0.400pt}{4.818pt}}
\put(539,41){\makebox(0,0){ }}
\put(539.0,539.0){\rule[-0.200pt]{0.400pt}{4.818pt}}
\put(70.0,82.0){\rule[-0.200pt]{0.400pt}{114.909pt}}
\put(70.0,82.0){\rule[-0.200pt]{112.982pt}{0.400pt}}
\put(539.0,82.0){\rule[-0.200pt]{0.400pt}{114.909pt}}
\put(70.0,559.0){\rule[-0.200pt]{112.982pt}{0.400pt}}
\put(148,190){\usebox{\plotpoint}}
\multiput(148.00,188.92)(0.817,-0.494){29}{\rule{0.750pt}{0.119pt}}
\multiput(148.00,189.17)(24.443,-16.000){2}{\rule{0.375pt}{0.400pt}}
\multiput(174.00,172.92)(2.043,-0.493){23}{\rule{1.700pt}{0.119pt}}
\multiput(174.00,173.17)(48.472,-13.000){2}{\rule{0.850pt}{0.400pt}}
\multiput(226.00,159.92)(1.091,-0.496){45}{\rule{0.967pt}{0.120pt}}
\multiput(226.00,160.17)(49.994,-24.000){2}{\rule{0.483pt}{0.400pt}}
\multiput(278.00,137.59)(11.619,0.477){7}{\rule{8.500pt}{0.115pt}}
\multiput(278.00,136.17)(87.358,5.000){2}{\rule{4.250pt}{0.400pt}}
\multiput(383.58,137.90)(0.497,-1.118){51}{\rule{0.120pt}{0.989pt}}
\multiput(382.17,139.95)(27.000,-57.948){2}{\rule{0.400pt}{0.494pt}}
\put(148,190){\makebox(0,0){$\blacklozenge$}}
\put(174,174){\makebox(0,0){$\blacklozenge$}}
\put(226,161){\makebox(0,0){$\blacklozenge$}}
\put(278,137){\makebox(0,0){$\blacklozenge$}}
\put(383,142){\makebox(0,0){$\blacklozenge$}}
\put(148,197){\usebox{\plotpoint}}
\multiput(148,197)(20.042,-5.396){2}{\usebox{\plotpoint}}
\multiput(174,190)(20.752,-0.399){2}{\usebox{\plotpoint}}
\multiput(226,189)(17.524,-11.121){3}{\usebox{\plotpoint}}
\multiput(278,156)(19.906,-5.877){6}{\usebox{\plotpoint}}
\multiput(383,125)(20.306,-4.296){2}{\usebox{\plotpoint}}
\multiput(435,114)(20.451,3.540){5}{\usebox{\plotpoint}}
\put(539,132){\usebox{\plotpoint}}
\put(148,197){\makebox(0,0){$\lozenge$}}
\put(174,190){\makebox(0,0){$\lozenge$}}
\put(226,189){\makebox(0,0){$\lozenge$}}
\put(278,156){\makebox(0,0){$\lozenge$}}
\put(383,125){\makebox(0,0){$\lozenge$}}
\put(435,114){\makebox(0,0){$\lozenge$}}
\put(539,132){\makebox(0,0){$\lozenge$}}
\sbox{\plotpoint}{\rule[-0.400pt]{0.800pt}{0.800pt}}%
\put(148,407){\usebox{\plotpoint}}
\multiput(148.00,408.40)(1.558,0.516){11}{\rule{2.511pt}{0.124pt}}
\multiput(148.00,405.34)(20.788,9.000){2}{\rule{1.256pt}{0.800pt}}
\multiput(174.00,417.41)(2.299,0.511){17}{\rule{3.667pt}{0.123pt}}
\multiput(174.00,414.34)(44.390,12.000){2}{\rule{1.833pt}{0.800pt}}
\multiput(226.00,426.08)(3.198,-0.516){11}{\rule{4.822pt}{0.124pt}}
\multiput(226.00,426.34)(41.991,-9.000){2}{\rule{2.411pt}{0.800pt}}
\multiput(278.00,417.09)(1.476,-0.503){65}{\rule{2.533pt}{0.121pt}}
\multiput(278.00,417.34)(99.742,-36.000){2}{\rule{1.267pt}{0.800pt}}
\multiput(384.41,383.00)(0.502,0.996){97}{\rule{0.121pt}{1.785pt}}
\multiput(381.34,383.00)(52.000,99.296){2}{\rule{0.800pt}{0.892pt}}
\multiput(435.00,487.41)(1.249,0.502){77}{\rule{2.181pt}{0.121pt}}
\multiput(435.00,484.34)(99.473,42.000){2}{\rule{1.090pt}{0.800pt}}
\put(148,407){\makebox(0,0){$\blacktriangledown$}}
\put(174,416){\makebox(0,0){$\blacktriangledown$}}
\put(226,428){\makebox(0,0){$\blacktriangledown$}}
\put(278,419){\makebox(0,0){$\blacktriangledown$}}
\put(383,383){\makebox(0,0){$\blacktriangledown$}}
\put(435,486){\makebox(0,0){$\blacktriangledown$}}
\put(539,528){\makebox(0,0){$\blacktriangledown$}}
\sbox{\plotpoint}{\rule[-0.200pt]{0.400pt}{0.400pt}}%
\put(70.0,82.0){\rule[-0.200pt]{0.400pt}{114.909pt}}
\put(70.0,82.0){\rule[-0.200pt]{112.982pt}{0.400pt}}
\put(539.0,82.0){\rule[-0.200pt]{0.400pt}{114.909pt}}
\put(70.0,559.0){\rule[-0.200pt]{112.982pt}{0.400pt}}
\end{picture}
&
      \hspace{0em}\begin{rotate}{90}\hspace{4.5em}{\small  length: b=30 bits}\end{rotate} \\ 
      \hspace{-6em}\begin{rotate}{90}\hspace{4.5em}{\small  length: b=100 bits}\end{rotate}& \hspace{-2.6em}
\setlength{\unitlength}{0.240900pt}
\ifx\plotpoint\undefined\newsavebox{\plotpoint}\fi
\begin{picture}(660,600)(0,0)
\sbox{\plotpoint}{\rule[-0.200pt]{0.400pt}{0.400pt}}%
\put(130.0,131.0){\rule[-0.200pt]{4.818pt}{0.400pt}}
\put(110,131){\makebox(0,0)[r]{0.01}}
\put(579.0,131.0){\rule[-0.200pt]{4.818pt}{0.400pt}}
\put(130.0,174.0){\rule[-0.200pt]{2.409pt}{0.400pt}}
\put(589.0,174.0){\rule[-0.200pt]{2.409pt}{0.400pt}}
\put(130.0,199.0){\rule[-0.200pt]{2.409pt}{0.400pt}}
\put(589.0,199.0){\rule[-0.200pt]{2.409pt}{0.400pt}}
\put(130.0,217.0){\rule[-0.200pt]{2.409pt}{0.400pt}}
\put(589.0,217.0){\rule[-0.200pt]{2.409pt}{0.400pt}}
\put(130.0,231.0){\rule[-0.200pt]{2.409pt}{0.400pt}}
\put(589.0,231.0){\rule[-0.200pt]{2.409pt}{0.400pt}}
\put(130.0,242.0){\rule[-0.200pt]{2.409pt}{0.400pt}}
\put(589.0,242.0){\rule[-0.200pt]{2.409pt}{0.400pt}}
\put(130.0,252.0){\rule[-0.200pt]{2.409pt}{0.400pt}}
\put(589.0,252.0){\rule[-0.200pt]{2.409pt}{0.400pt}}
\put(130.0,260.0){\rule[-0.200pt]{2.409pt}{0.400pt}}
\put(589.0,260.0){\rule[-0.200pt]{2.409pt}{0.400pt}}
\put(130.0,267.0){\rule[-0.200pt]{2.409pt}{0.400pt}}
\put(589.0,267.0){\rule[-0.200pt]{2.409pt}{0.400pt}}
\put(130.0,274.0){\rule[-0.200pt]{4.818pt}{0.400pt}}
\put(110,274){\makebox(0,0)[r]{0.1}}
\put(579.0,274.0){\rule[-0.200pt]{4.818pt}{0.400pt}}
\put(130.0,317.0){\rule[-0.200pt]{2.409pt}{0.400pt}}
\put(589.0,317.0){\rule[-0.200pt]{2.409pt}{0.400pt}}
\put(130.0,342.0){\rule[-0.200pt]{2.409pt}{0.400pt}}
\put(589.0,342.0){\rule[-0.200pt]{2.409pt}{0.400pt}}
\put(130.0,360.0){\rule[-0.200pt]{2.409pt}{0.400pt}}
\put(589.0,360.0){\rule[-0.200pt]{2.409pt}{0.400pt}}
\put(130.0,373.0){\rule[-0.200pt]{2.409pt}{0.400pt}}
\put(589.0,373.0){\rule[-0.200pt]{2.409pt}{0.400pt}}
\put(130.0,385.0){\rule[-0.200pt]{2.409pt}{0.400pt}}
\put(589.0,385.0){\rule[-0.200pt]{2.409pt}{0.400pt}}
\put(130.0,394.0){\rule[-0.200pt]{2.409pt}{0.400pt}}
\put(589.0,394.0){\rule[-0.200pt]{2.409pt}{0.400pt}}
\put(130.0,403.0){\rule[-0.200pt]{2.409pt}{0.400pt}}
\put(589.0,403.0){\rule[-0.200pt]{2.409pt}{0.400pt}}
\put(130.0,410.0){\rule[-0.200pt]{2.409pt}{0.400pt}}
\put(589.0,410.0){\rule[-0.200pt]{2.409pt}{0.400pt}}
\put(130.0,416.0){\rule[-0.200pt]{4.818pt}{0.400pt}}
\put(110,416){\makebox(0,0)[r]{1}}
\put(579.0,416.0){\rule[-0.200pt]{4.818pt}{0.400pt}}
\put(130.0,459.0){\rule[-0.200pt]{2.409pt}{0.400pt}}
\put(589.0,459.0){\rule[-0.200pt]{2.409pt}{0.400pt}}
\put(130.0,484.0){\rule[-0.200pt]{2.409pt}{0.400pt}}
\put(589.0,484.0){\rule[-0.200pt]{2.409pt}{0.400pt}}
\put(130.0,502.0){\rule[-0.200pt]{2.409pt}{0.400pt}}
\put(589.0,502.0){\rule[-0.200pt]{2.409pt}{0.400pt}}
\put(130.0,516.0){\rule[-0.200pt]{2.409pt}{0.400pt}}
\put(589.0,516.0){\rule[-0.200pt]{2.409pt}{0.400pt}}
\put(130.0,527.0){\rule[-0.200pt]{2.409pt}{0.400pt}}
\put(589.0,527.0){\rule[-0.200pt]{2.409pt}{0.400pt}}
\put(130.0,537.0){\rule[-0.200pt]{2.409pt}{0.400pt}}
\put(589.0,537.0){\rule[-0.200pt]{2.409pt}{0.400pt}}
\put(130.0,545.0){\rule[-0.200pt]{2.409pt}{0.400pt}}
\put(589.0,545.0){\rule[-0.200pt]{2.409pt}{0.400pt}}
\put(130.0,552.0){\rule[-0.200pt]{2.409pt}{0.400pt}}
\put(589.0,552.0){\rule[-0.200pt]{2.409pt}{0.400pt}}
\put(130.0,559.0){\rule[-0.200pt]{4.818pt}{0.400pt}}
\put(110,559){\makebox(0,0)[r]{10}}
\put(579.0,559.0){\rule[-0.200pt]{4.818pt}{0.400pt}}
\put(130.0,131.0){\rule[-0.200pt]{0.400pt}{4.818pt}}
\put(130,90){\makebox(0,0){1}}
\put(130.0,539.0){\rule[-0.200pt]{0.400pt}{4.818pt}}
\put(182.0,131.0){\rule[-0.200pt]{0.400pt}{4.818pt}}
\put(182,90){\makebox(0,0){2}}
\put(182.0,539.0){\rule[-0.200pt]{0.400pt}{4.818pt}}
\put(234.0,131.0){\rule[-0.200pt]{0.400pt}{4.818pt}}
\put(234,90){\makebox(0,0){3}}
\put(234.0,539.0){\rule[-0.200pt]{0.400pt}{4.818pt}}
\put(286.0,131.0){\rule[-0.200pt]{0.400pt}{4.818pt}}
\put(286,90){\makebox(0,0){4}}
\put(286.0,539.0){\rule[-0.200pt]{0.400pt}{4.818pt}}
\put(338.0,131.0){\rule[-0.200pt]{0.400pt}{4.818pt}}
\put(338,90){\makebox(0,0){5}}
\put(338.0,539.0){\rule[-0.200pt]{0.400pt}{4.818pt}}
\put(391.0,131.0){\rule[-0.200pt]{0.400pt}{4.818pt}}
\put(391,90){\makebox(0,0){6}}
\put(391.0,539.0){\rule[-0.200pt]{0.400pt}{4.818pt}}
\put(443.0,131.0){\rule[-0.200pt]{0.400pt}{4.818pt}}
\put(443,90){\makebox(0,0){7}}
\put(443.0,539.0){\rule[-0.200pt]{0.400pt}{4.818pt}}
\put(495.0,131.0){\rule[-0.200pt]{0.400pt}{4.818pt}}
\put(495,90){\makebox(0,0){8}}
\put(495.0,539.0){\rule[-0.200pt]{0.400pt}{4.818pt}}
\put(547.0,131.0){\rule[-0.200pt]{0.400pt}{4.818pt}}
\put(547,90){\makebox(0,0){9}}
\put(547.0,539.0){\rule[-0.200pt]{0.400pt}{4.818pt}}
\put(599.0,131.0){\rule[-0.200pt]{0.400pt}{4.818pt}}
\put(599,90){\makebox(0,0){10}}
\put(599.0,539.0){\rule[-0.200pt]{0.400pt}{4.818pt}}
\put(130.0,131.0){\rule[-0.200pt]{0.400pt}{103.105pt}}
\put(130.0,131.0){\rule[-0.200pt]{112.982pt}{0.400pt}}
\put(599.0,131.0){\rule[-0.200pt]{0.400pt}{103.105pt}}
\put(130.0,559.0){\rule[-0.200pt]{112.982pt}{0.400pt}}
\put(364,29){\makebox(0,0){$B/b$}}
\put(208,266){\usebox{\plotpoint}}
\multiput(208.00,264.92)(0.873,-0.494){27}{\rule{0.793pt}{0.119pt}}
\multiput(208.00,265.17)(24.353,-15.000){2}{\rule{0.397pt}{0.400pt}}
\multiput(234.00,249.94)(7.500,-0.468){5}{\rule{5.300pt}{0.113pt}}
\multiput(234.00,250.17)(41.000,-4.000){2}{\rule{2.650pt}{0.400pt}}
\multiput(286.00,245.92)(0.744,-0.498){67}{\rule{0.694pt}{0.120pt}}
\multiput(286.00,246.17)(50.559,-35.000){2}{\rule{0.347pt}{0.400pt}}
\multiput(338.00,210.92)(3.575,-0.494){27}{\rule{2.900pt}{0.119pt}}
\multiput(338.00,211.17)(98.981,-15.000){2}{\rule{1.450pt}{0.400pt}}
\multiput(443.00,195.92)(0.591,-0.498){85}{\rule{0.573pt}{0.120pt}}
\multiput(443.00,196.17)(50.811,-44.000){2}{\rule{0.286pt}{0.400pt}}
\multiput(495.00,151.92)(1.145,-0.496){41}{\rule{1.009pt}{0.120pt}}
\multiput(495.00,152.17)(47.906,-22.000){2}{\rule{0.505pt}{0.400pt}}
\put(208,266){\makebox(0,0){$\blacklozenge$}}
\put(234,251){\makebox(0,0){$\blacklozenge$}}
\put(286,247){\makebox(0,0){$\blacklozenge$}}
\put(338,212){\makebox(0,0){$\blacklozenge$}}
\put(443,197){\makebox(0,0){$\blacklozenge$}}
\put(495,153){\makebox(0,0){$\blacklozenge$}}
\put(208,274){\usebox{\plotpoint}}
\multiput(208,274)(19.838,-6.104){2}{\usebox{\plotpoint}}
\multiput(234,266)(20.382,-3.920){2}{\usebox{\plotpoint}}
\multiput(286,256)(19.614,-6.789){3}{\usebox{\plotpoint}}
\multiput(338,238)(19.457,-7.227){5}{\usebox{\plotpoint}}
\multiput(443,199)(19.614,-6.789){3}{\usebox{\plotpoint}}
\multiput(495,181)(19.891,-5.929){5}{\usebox{\plotpoint}}
\put(599,150){\usebox{\plotpoint}}
\put(208,274){\makebox(0,0){$\lozenge$}}
\put(234,266){\makebox(0,0){$\lozenge$}}
\put(286,256){\makebox(0,0){$\lozenge$}}
\put(338,238){\makebox(0,0){$\lozenge$}}
\put(443,199){\makebox(0,0){$\lozenge$}}
\put(495,181){\makebox(0,0){$\lozenge$}}
\put(599,150){\makebox(0,0){$\lozenge$}}
\sbox{\plotpoint}{\rule[-0.400pt]{0.800pt}{0.800pt}}%
\put(208,425){\usebox{\plotpoint}}
\multiput(208.00,426.40)(2.139,0.526){7}{\rule{3.171pt}{0.127pt}}
\multiput(208.00,423.34)(19.418,7.000){2}{\rule{1.586pt}{0.800pt}}
\multiput(234.00,430.08)(4.417,-0.526){7}{\rule{6.143pt}{0.127pt}}
\multiput(234.00,430.34)(39.250,-7.000){2}{\rule{3.071pt}{0.800pt}}
\multiput(286.00,426.41)(1.577,0.507){27}{\rule{2.647pt}{0.122pt}}
\multiput(286.00,423.34)(46.506,17.000){2}{\rule{1.324pt}{0.800pt}}
\multiput(338.00,440.09)(2.241,-0.504){41}{\rule{3.700pt}{0.122pt}}
\multiput(338.00,440.34)(97.320,-24.000){2}{\rule{1.850pt}{0.800pt}}
\multiput(443.00,419.41)(1.012,0.504){45}{\rule{1.800pt}{0.121pt}}
\multiput(443.00,416.34)(48.264,26.000){2}{\rule{0.900pt}{0.800pt}}
\multiput(495.00,445.41)(3.643,0.508){23}{\rule{5.747pt}{0.122pt}}
\multiput(495.00,442.34)(92.073,15.000){2}{\rule{2.873pt}{0.800pt}}
\put(208,425){\makebox(0,0){$\blacktriangledown$}}
\put(234,432){\makebox(0,0){$\blacktriangledown$}}
\put(286,425){\makebox(0,0){$\blacktriangledown$}}
\put(338,442){\makebox(0,0){$\blacktriangledown$}}
\put(443,418){\makebox(0,0){$\blacktriangledown$}}
\put(495,444){\makebox(0,0){$\blacktriangledown$}}
\put(599,459){\makebox(0,0){$\blacktriangledown$}}
\sbox{\plotpoint}{\rule[-0.200pt]{0.400pt}{0.400pt}}%
\put(130.0,131.0){\rule[-0.200pt]{0.400pt}{103.105pt}}
\put(130.0,131.0){\rule[-0.200pt]{112.982pt}{0.400pt}}
\put(599.0,131.0){\rule[-0.200pt]{0.400pt}{103.105pt}}
\put(130.0,559.0){\rule[-0.200pt]{112.982pt}{0.400pt}}
\end{picture} & %
\hspace{-2em}
\setlength{\unitlength}{0.240900pt}
\ifx\plotpoint\undefined\newsavebox{\plotpoint}\fi
\begin{picture}(600,600)(0,0)
\sbox{\plotpoint}{\rule[-0.200pt]{0.400pt}{0.400pt}}%
\put(70.0,131.0){\rule[-0.200pt]{4.818pt}{0.400pt}}
\put(50,131){\makebox(0,0)[r]{ }}
\put(519.0,131.0){\rule[-0.200pt]{4.818pt}{0.400pt}}
\put(70.0,174.0){\rule[-0.200pt]{2.409pt}{0.400pt}}
\put(529.0,174.0){\rule[-0.200pt]{2.409pt}{0.400pt}}
\put(70.0,199.0){\rule[-0.200pt]{2.409pt}{0.400pt}}
\put(529.0,199.0){\rule[-0.200pt]{2.409pt}{0.400pt}}
\put(70.0,217.0){\rule[-0.200pt]{2.409pt}{0.400pt}}
\put(529.0,217.0){\rule[-0.200pt]{2.409pt}{0.400pt}}
\put(70.0,231.0){\rule[-0.200pt]{2.409pt}{0.400pt}}
\put(529.0,231.0){\rule[-0.200pt]{2.409pt}{0.400pt}}
\put(70.0,242.0){\rule[-0.200pt]{2.409pt}{0.400pt}}
\put(529.0,242.0){\rule[-0.200pt]{2.409pt}{0.400pt}}
\put(70.0,252.0){\rule[-0.200pt]{2.409pt}{0.400pt}}
\put(529.0,252.0){\rule[-0.200pt]{2.409pt}{0.400pt}}
\put(70.0,260.0){\rule[-0.200pt]{2.409pt}{0.400pt}}
\put(529.0,260.0){\rule[-0.200pt]{2.409pt}{0.400pt}}
\put(70.0,267.0){\rule[-0.200pt]{2.409pt}{0.400pt}}
\put(529.0,267.0){\rule[-0.200pt]{2.409pt}{0.400pt}}
\put(70.0,274.0){\rule[-0.200pt]{4.818pt}{0.400pt}}
\put(50,274){\makebox(0,0)[r]{ }}
\put(519.0,274.0){\rule[-0.200pt]{4.818pt}{0.400pt}}
\put(70.0,317.0){\rule[-0.200pt]{2.409pt}{0.400pt}}
\put(529.0,317.0){\rule[-0.200pt]{2.409pt}{0.400pt}}
\put(70.0,342.0){\rule[-0.200pt]{2.409pt}{0.400pt}}
\put(529.0,342.0){\rule[-0.200pt]{2.409pt}{0.400pt}}
\put(70.0,360.0){\rule[-0.200pt]{2.409pt}{0.400pt}}
\put(529.0,360.0){\rule[-0.200pt]{2.409pt}{0.400pt}}
\put(70.0,373.0){\rule[-0.200pt]{2.409pt}{0.400pt}}
\put(529.0,373.0){\rule[-0.200pt]{2.409pt}{0.400pt}}
\put(70.0,385.0){\rule[-0.200pt]{2.409pt}{0.400pt}}
\put(529.0,385.0){\rule[-0.200pt]{2.409pt}{0.400pt}}
\put(70.0,394.0){\rule[-0.200pt]{2.409pt}{0.400pt}}
\put(529.0,394.0){\rule[-0.200pt]{2.409pt}{0.400pt}}
\put(70.0,403.0){\rule[-0.200pt]{2.409pt}{0.400pt}}
\put(529.0,403.0){\rule[-0.200pt]{2.409pt}{0.400pt}}
\put(70.0,410.0){\rule[-0.200pt]{2.409pt}{0.400pt}}
\put(529.0,410.0){\rule[-0.200pt]{2.409pt}{0.400pt}}
\put(70.0,416.0){\rule[-0.200pt]{4.818pt}{0.400pt}}
\put(50,416){\makebox(0,0)[r]{ }}
\put(519.0,416.0){\rule[-0.200pt]{4.818pt}{0.400pt}}
\put(70.0,459.0){\rule[-0.200pt]{2.409pt}{0.400pt}}
\put(529.0,459.0){\rule[-0.200pt]{2.409pt}{0.400pt}}
\put(70.0,484.0){\rule[-0.200pt]{2.409pt}{0.400pt}}
\put(529.0,484.0){\rule[-0.200pt]{2.409pt}{0.400pt}}
\put(70.0,502.0){\rule[-0.200pt]{2.409pt}{0.400pt}}
\put(529.0,502.0){\rule[-0.200pt]{2.409pt}{0.400pt}}
\put(70.0,516.0){\rule[-0.200pt]{2.409pt}{0.400pt}}
\put(529.0,516.0){\rule[-0.200pt]{2.409pt}{0.400pt}}
\put(70.0,527.0){\rule[-0.200pt]{2.409pt}{0.400pt}}
\put(529.0,527.0){\rule[-0.200pt]{2.409pt}{0.400pt}}
\put(70.0,537.0){\rule[-0.200pt]{2.409pt}{0.400pt}}
\put(529.0,537.0){\rule[-0.200pt]{2.409pt}{0.400pt}}
\put(70.0,545.0){\rule[-0.200pt]{2.409pt}{0.400pt}}
\put(529.0,545.0){\rule[-0.200pt]{2.409pt}{0.400pt}}
\put(70.0,552.0){\rule[-0.200pt]{2.409pt}{0.400pt}}
\put(529.0,552.0){\rule[-0.200pt]{2.409pt}{0.400pt}}
\put(70.0,559.0){\rule[-0.200pt]{4.818pt}{0.400pt}}
\put(50,559){\makebox(0,0)[r]{ }}
\put(519.0,559.0){\rule[-0.200pt]{4.818pt}{0.400pt}}
\put(70.0,131.0){\rule[-0.200pt]{0.400pt}{4.818pt}}
\put(70,90){\makebox(0,0){1}}
\put(70.0,539.0){\rule[-0.200pt]{0.400pt}{4.818pt}}
\put(122.0,131.0){\rule[-0.200pt]{0.400pt}{4.818pt}}
\put(122,90){\makebox(0,0){2}}
\put(122.0,539.0){\rule[-0.200pt]{0.400pt}{4.818pt}}
\put(174.0,131.0){\rule[-0.200pt]{0.400pt}{4.818pt}}
\put(174,90){\makebox(0,0){3}}
\put(174.0,539.0){\rule[-0.200pt]{0.400pt}{4.818pt}}
\put(226.0,131.0){\rule[-0.200pt]{0.400pt}{4.818pt}}
\put(226,90){\makebox(0,0){4}}
\put(226.0,539.0){\rule[-0.200pt]{0.400pt}{4.818pt}}
\put(278.0,131.0){\rule[-0.200pt]{0.400pt}{4.818pt}}
\put(278,90){\makebox(0,0){5}}
\put(278.0,539.0){\rule[-0.200pt]{0.400pt}{4.818pt}}
\put(331.0,131.0){\rule[-0.200pt]{0.400pt}{4.818pt}}
\put(331,90){\makebox(0,0){6}}
\put(331.0,539.0){\rule[-0.200pt]{0.400pt}{4.818pt}}
\put(383.0,131.0){\rule[-0.200pt]{0.400pt}{4.818pt}}
\put(383,90){\makebox(0,0){7}}
\put(383.0,539.0){\rule[-0.200pt]{0.400pt}{4.818pt}}
\put(435.0,131.0){\rule[-0.200pt]{0.400pt}{4.818pt}}
\put(435,90){\makebox(0,0){8}}
\put(435.0,539.0){\rule[-0.200pt]{0.400pt}{4.818pt}}
\put(487.0,131.0){\rule[-0.200pt]{0.400pt}{4.818pt}}
\put(487,90){\makebox(0,0){9}}
\put(487.0,539.0){\rule[-0.200pt]{0.400pt}{4.818pt}}
\put(539.0,131.0){\rule[-0.200pt]{0.400pt}{4.818pt}}
\put(539,90){\makebox(0,0){10}}
\put(539.0,539.0){\rule[-0.200pt]{0.400pt}{4.818pt}}
\put(70.0,131.0){\rule[-0.200pt]{0.400pt}{103.105pt}}
\put(70.0,131.0){\rule[-0.200pt]{112.982pt}{0.400pt}}
\put(539.0,131.0){\rule[-0.200pt]{0.400pt}{103.105pt}}
\put(70.0,559.0){\rule[-0.200pt]{112.982pt}{0.400pt}}
\put(304,29){\makebox(0,0){$B/b$}}
\put(148,318){\usebox{\plotpoint}}
\multiput(148.00,316.92)(1.012,-0.493){23}{\rule{0.900pt}{0.119pt}}
\multiput(148.00,317.17)(24.132,-13.000){2}{\rule{0.450pt}{0.400pt}}
\multiput(174.00,303.93)(2.999,-0.489){15}{\rule{2.411pt}{0.118pt}}
\multiput(174.00,304.17)(46.996,-9.000){2}{\rule{1.206pt}{0.400pt}}
\multiput(226.00,294.92)(0.968,-0.497){51}{\rule{0.870pt}{0.120pt}}
\multiput(226.00,295.17)(50.194,-27.000){2}{\rule{0.435pt}{0.400pt}}
\multiput(278.00,267.92)(2.663,-0.496){37}{\rule{2.200pt}{0.119pt}}
\multiput(278.00,268.17)(100.434,-20.000){2}{\rule{1.100pt}{0.400pt}}
\multiput(383.00,247.92)(0.723,-0.498){69}{\rule{0.678pt}{0.120pt}}
\multiput(383.00,248.17)(50.593,-36.000){2}{\rule{0.339pt}{0.400pt}}
\multiput(435.00,211.92)(1.160,-0.498){87}{\rule{1.024pt}{0.120pt}}
\multiput(435.00,212.17)(101.874,-45.000){2}{\rule{0.512pt}{0.400pt}}
\put(148,318){\makebox(0,0){$\blacklozenge$}}
\put(174,305){\makebox(0,0){$\blacklozenge$}}
\put(226,296){\makebox(0,0){$\blacklozenge$}}
\put(278,269){\makebox(0,0){$\blacklozenge$}}
\put(383,249){\makebox(0,0){$\blacklozenge$}}
\put(435,213){\makebox(0,0){$\blacklozenge$}}
\put(539,168){\makebox(0,0){$\blacklozenge$}}
\put(148,326){\usebox{\plotpoint}}
\multiput(148,326)(18.564,-9.282){2}{\usebox{\plotpoint}}
\multiput(174,313)(20.451,-3.540){2}{\usebox{\plotpoint}}
\multiput(226,304)(19.115,-8.087){3}{\usebox{\plotpoint}}
\multiput(278,282)(20.489,-3.317){5}{\usebox{\plotpoint}}
\multiput(383,265)(17.828,-10.628){3}{\usebox{\plotpoint}}
\multiput(435,234)(19.671,-6.620){5}{\usebox{\plotpoint}}
\put(539,199){\usebox{\plotpoint}}
\put(148,326){\makebox(0,0){$\lozenge$}}
\put(174,313){\makebox(0,0){$\lozenge$}}
\put(226,304){\makebox(0,0){$\lozenge$}}
\put(278,282){\makebox(0,0){$\lozenge$}}
\put(383,265){\makebox(0,0){$\lozenge$}}
\put(435,234){\makebox(0,0){$\lozenge$}}
\put(539,199){\makebox(0,0){$\lozenge$}}
\sbox{\plotpoint}{\rule[-0.400pt]{0.800pt}{0.800pt}}%
\put(148,425){\usebox{\plotpoint}}
\put(148,422.84){\rule{6.263pt}{0.800pt}}
\multiput(148.00,423.34)(13.000,-1.000){2}{\rule{3.132pt}{0.800pt}}
\put(174,422.84){\rule{12.527pt}{0.800pt}}
\multiput(174.00,422.34)(26.000,1.000){2}{\rule{6.263pt}{0.800pt}}
\put(226,425.34){\rule{10.600pt}{0.800pt}}
\multiput(226.00,423.34)(29.999,4.000){2}{\rule{5.300pt}{0.800pt}}
\put(278,429.34){\rule{21.200pt}{0.800pt}}
\multiput(278.00,427.34)(60.998,4.000){2}{\rule{10.600pt}{0.800pt}}
\put(383,433.34){\rule{10.600pt}{0.800pt}}
\multiput(383.00,431.34)(29.999,4.000){2}{\rule{5.300pt}{0.800pt}}
\multiput(435.00,438.40)(5.719,0.514){13}{\rule{8.520pt}{0.124pt}}
\multiput(435.00,435.34)(86.316,10.000){2}{\rule{4.260pt}{0.800pt}}
\put(148,425){\makebox(0,0){$\blacktriangledown$}}
\put(174,424){\makebox(0,0){$\blacktriangledown$}}
\put(226,425){\makebox(0,0){$\blacktriangledown$}}
\put(278,429){\makebox(0,0){$\blacktriangledown$}}
\put(383,433){\makebox(0,0){$\blacktriangledown$}}
\put(435,437){\makebox(0,0){$\blacktriangledown$}}
\put(539,447){\makebox(0,0){$\blacktriangledown$}}
\sbox{\plotpoint}{\rule[-0.200pt]{0.400pt}{0.400pt}}%
\put(70.0,131.0){\rule[-0.200pt]{0.400pt}{103.105pt}}
\put(70.0,131.0){\rule[-0.200pt]{112.982pt}{0.400pt}}
\put(539.0,131.0){\rule[-0.200pt]{0.400pt}{103.105pt}}
\put(70.0,559.0){\rule[-0.200pt]{112.982pt}{0.400pt}}
\end{picture}
&
      \hspace{0em}\begin{rotate}{90}\hspace{4.5em}{\small  length: b=300 bits}\end{rotate} \\ 
      \multicolumn{3}{c}{\hspace{-2em}\setlength{\unitlength}{1em}
\begin{picture}(30,3)(0,0)
  \put(0,2){\line(1,0){4}}
  \put(2,2){\makebox(0,0){$\blacklozenge$}}
  %
  \multiput(10,2)(0.4,0){10}{\circle*{0.000001}}
  \put(12,2){\makebox(0,0){$\lozenge$}}
  \put(20,2){\line(1,0){4}}
  \put(22,2){\makebox(0,0){$\blacktriangledown$}}
  \put(2,2.5){\makebox(0,0)[b]{$A=\rho(c(q(h,r),h))$}}
  \put(12,2.5){\makebox(0,0)[b]{$B=\rho(c(b(h,r),h))$}}
  \put(22,2.5){\makebox(0,0)[b]{$B/A$}}
\end{picture}}
    \end{tabular}
  \end{center}

  \vspace{-3em}
  
  \caption{Correlation of the encrypted hidden content with the
    original hidden content when the randomization has been obtained
    with insertion of random bits in fixed positions (line with
    $\lozenge$) and with the method presented here (line with
    $\blacklozenge$).}
  \label{correl2}
\end{figure*}
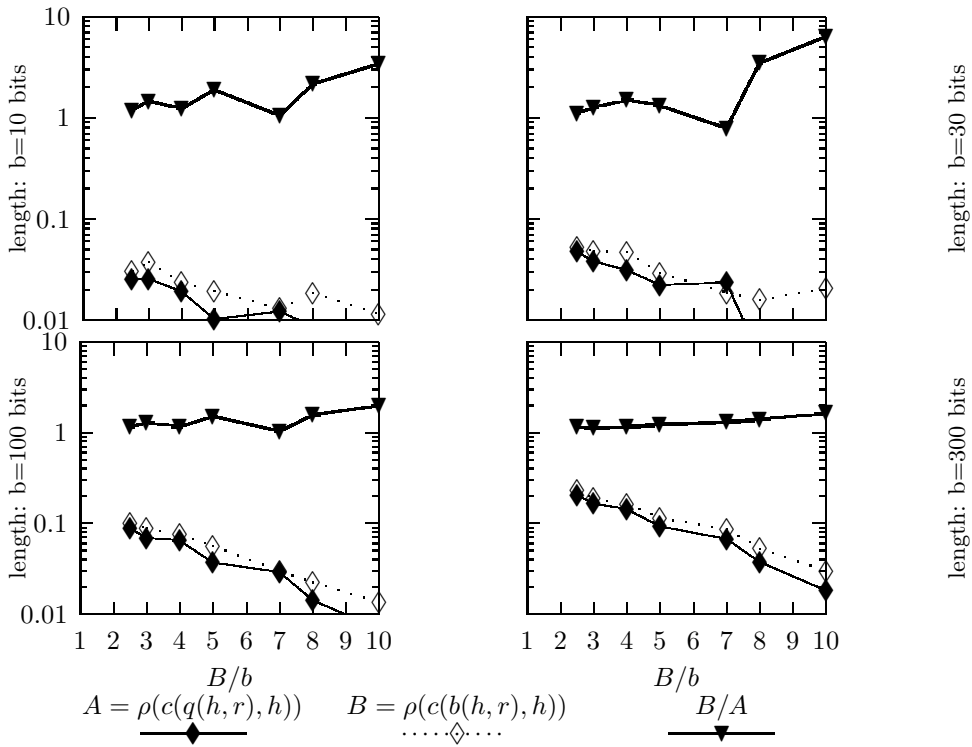

Let $b(h,r)$ be the function that randomizes $h$ by adding random bits
in fixed positions, and $q(h,r)$ the function that randomizes $h$
using the sequence $p_n$. We have seen that $\rho(b(h,r),h)=1$, while
$\rho(q(h,r),h)=1$ is given in Figure~\ref{correl1}. The use of
one-time-notepad reduces considerably the correlation with $h$, but in
different measure for the two randomization methods
(Figure~\ref{correl2}).
That is, even after encryption, adding random bits in fixed positions
is more vulnerable to correlation attacks that the randomization
method we are using here.  The encryption is based on a symmetric
method and a secure key distribution protocol. We have proved that the
protocol is secure under the hypothesis of the Diffie-Hellman decision
problem. Note, however, that the protocol as we have presented it is
vulnerable to a man-in-the-middle attack \cite{desmedt:11}: an
intruder, listening to the traffic and posing as one of the members of
the group would obtain the key. Resisting this kind of attacks
requires the introduction of an authentication protocol
\cite{simmons:88}, which can be done using standard methods and which
goes beyond the scope of this paper.

As an additional issue, if we want to protect our data against the
owner of the social network on which we post, we can't obviously rely
on any means that the owner itself places at our disposal. This means
that there has to be a way to encrypt our tags \emph{despite} the
owner. Doing this entails system and legal issue that are, also,
beyond the scope of this paper.

\end{document}